\documentclass[12pt,a4paper]{article}

\usepackage{amsmath,amssymb}
\usepackage{mathrsfs}

\newcommand{\be}{\begin{equation}}
\newcommand{\ee}{\end{equation}}

\renewcommand{\theequation}{\arabic{section}.\arabic{equation}}

\def\Tr{{\rm Tr}\,}
\def\cN{{\cal N}}

\def\bea{\begin{eqnarray}}
\def\eea{\end{eqnarray}}
\def\nn{\nonumber}

\def\theequation{\arabic{section}.\arabic{equation}}

\begin{document}
\begin{titlepage}

\phantom{1}

\vspace{2cm}

\begin{center}
{\Large\bf Low-energy effective action in two-dimensional
\\[2mm]
SQED: A two-loop analysis} \vspace{1cm}

{\large\bf I.B. Samsonov
\\[8pt]
\it\small  Bogoliubov Laboratory of Theoretical Physics, JINR,
141980 Dubna, Moscow region, Russia}
\\
{\tt Email:\ samsonov@theor.jinr.ru}
\end{center}
\vspace{0.5cm}

\begin{abstract}
We study two-loop quantum corrections to the low-energy effective
actions in $\cN=(2,2)$ and $\cN=(4,4)$ SQED in the Coulomb branch.
In the latter model, the low-energy effective action is described
by a generalized K\"ahler potential which depends on both chiral
and twisted chiral superfields. We demonstrate that this
generalized K\"ahler potential is one-loop exact and corresponds
to the $\cN=(4,4)$ sigma-model with torsion presented by Ro\v cek,
Schoutens and Sevrin \cite{Rocek91}. In the $\cN=(2,2)$ SQED, the
effective K\"ahler potential is not protected against higher-loop
quantum corrections. The two-loop quantum corrections to this
potential and the corresponding sigma-model metric are explicitly
found.
\end{abstract}

\end{titlepage}

\tableofcontents

\setcounter{equation}{0}
\section{Introduction and summary}

Two-dimensional supersymmetric gauge theories have a wide range of
applications in physics and geometry. In field theory, 2d
gauged linear sigma-models in $\cN=(2,2)$ superspace serve as
canonical examples which provide very useful insights on
low-energy dynamics of four-dimensional supersymmetric gauge
theories \cite{DAdda,Witten93}. Geometrically,
two-dimensional non-linear sigma-models with extended
supersymmetry appear very reach because of existence of numerous
twisted-chiral multiplets \cite{GHR} which possess no analogs in
higher dimensions. In particular, such sigma-models may support K\"ahler,
hyper-K\"ahler or generalized K\"ahler geometry. It is natural to expect that
some of these geometries may arise as
low-energy effective actions in two-dimensional gauge theories in
$\cN=(2,2)$ superspace.

The study of low-energy effective action in Abelian gauge theories
in $\cN=(2,2)$ superspace was initiated long ago \cite{DAdda}. The authors of
this work showed that the field strength of $\cN=(2,2)$ vector
multiplet is given by a twisted chiral superfield which we denote by $\Sigma$
throughout this work. The effective
action for $\Sigma$ may have a superpotential $W(\Sigma)$
and a K\"ahler potential $K(\Sigma,\bar\Sigma)$, as well as higher-derivative terms which form together
the Euler-Heisenberg-type effective action in two-dimensional SQED.
The structure of one-loop quantum corrections to these potentials was found in
\cite{DAdda}:
\bea
W^{(1)}(\Sigma) &\propto& \Sigma\ln \Sigma - \Sigma\,,
\label1\\
K^{(1)}(\Sigma,\bar\Sigma) &\propto& \ln \Sigma \ln \bar\Sigma\,.
\label2
\eea
At leading order, one can discard higher-derivative terms in the effective action
and treat the low-energy theory as a $(2,2)$ sigma-model with
the K\"ahler potential (\ref2) and superpotential (\ref1).

The superpotential (\ref1) is known to be one-loop exact and its form is completely determined by the anomaly
of $U(1)\times U(1)$ R-symmetry \cite{DAdda}. The K\"ahler potential can,
however, receive higher-loop quantum corrections. This paper aims
to trigger the study of quantum corrections to the effective
K\"ahler potential $K(\Sigma,\bar\Sigma)$ and corresponding sigma-model geometry
{\it beyond one-loop order}.

We consider two-loop effective action in $\cN=(2,2)$ and $(4,4)$ SQED in
the Coulomb branch. In the $(2,2)$ case, the Coulomb branch is known
to exist only when the $U(1)$ charges of chiral multiplets sum to
zero \cite{Witten93,Hori}. This is typically satisfied for the SQED with two
chiral multiplets which carry opposite charges with respect to the
gauge group. For this theory we explicitly compute two-loop
quantum corrections to the effective K\"ahler potential $K^{(2)}(\Sigma,\bar\Sigma)$.

An important feature of two-dimensional gauge
theories is that Feynman graphs with internal (super)photon lines
suffer from IR divergencies. We show that for supersymmetric
gauge theories in the $\cN=(2,2)$ superspace it is possible to
introduce gauge invariant mass term for the vector multiplet which
naturally regulates such IR divergencies. This mass term may be
obtained by the dimensional reduction from the three-dimensional
(super) Chern-Simons action which is also known to be responsible
for the gauge-invariant mass of the vector multiplet in three dimensions. In our case, the
two-loop quantum corrections to the effective action explicitly
depend on the vector multiplet mass and are singular in the limit
when this mass vanishes.

The $(4,4)$ vector multiplet in the $\cN=(2,2)$ superspace is
described by a pair of chiral $\Phi$ and twisted chiral $\Sigma$ multiplets
\cite{GHR}. At leading order in the derivative expansion,
the low-energy effective action in the $\cN=(4,4)$
SQED is described by a generalized K\"ahler potential
$K(\Sigma,\bar\Sigma;\Phi,\bar\Phi)$. Performing
explicit quantum computations we demonstrate that this potential
does not receive two-loop quantum corrections and is
one-loop exact. At one-loop order, this function coincides with
the potential for the $(4,4)$ sigma-model with torsion studied in
\cite{Rocek91}\footnote{This sigma-model can be considered as a
particular case of the $\cN=(4,4)$ super-Liouville theory which was
constructed originally in \cite{IK,IKL}. I am grateful to E.A. Ivanov for drawing
my attention to these works.}
\be
K^{(1)}(\Sigma,\bar\Sigma;\Phi,\bar\Phi)
=\frac1{4\pi} \left[
\ln \Sigma \ln \bar\Sigma + {\rm Li}_2\left(
-\frac{\Phi\bar\Phi}{\Sigma\bar\Sigma}
\right)
\right].
\label{3}
\ee
This sigma-model is known to contain the
Wess-Zumino term which has rigid form because of its
topological nature \cite{GHR,semichiral1}. The coefficient in front of the Wess-Zumino
term is one-loop exact and  quantizes (see, e.g.,
\cite{Bardeen}). This confirms the non-renormalization of the
potential (\ref{3}) claimed in \cite{Seib97}.

Qualitatively, the presence of the Wess-Zumino term in the
low-energy effective action of $\cN=(4,4)$ SQED is well understood.
Indeed, this theory possesses $SO(4)\simeq SU(2)\times SU(2)$ R-symmetry which
suffers from the 't~Hooft anomaly. At low energy, we integrate out the massive chiral
multiplets and consider effective action for the light vector
multiplet. However, the total contribution to the anomaly should
be the same at low and high energies since the anomaly cannot
depend on the energy scale. Thus, the low-energy effective action must
include the Wess-Zumino term  compensating the contribution
to the anomaly from the fermions that were integrated out. This
statement is well known as the 't~Hooft anomaly matching argument
\cite{tHooft}.

It is pertinent to mention here the amazing analogy of the effective
potentials (\ref{1}), (\ref{2}) and (\ref{3}) with certain terms in low-energy
effective actions of {\it four-dimensional} $\cN=2$ and $\cN=4$ gauge
theories. Recall that the 4d $\cN=2$ gauge multiplet may be described
by an $\cN=2$ chiral superfield $\cal W$. The superpotential
(\ref{1}) is somewhat similar to the so-called holomorphic potential
\cite{Seiberg88} ${\cal F}({\cal W})\propto {\cal W}^2 \ln {\cal W}$ while
the K\"ahler potential (\ref{2}) formally coincides with the non-holomorphic
potential \cite{DS} ${\cal H}({\cal W},\bar{\cal W})\propto \ln{\cal W}\ln\bar{\cal
W}$. This analogy is not accidental: both ${\cal F}({\cal W})$
and the superpotential (\ref{1}) appear as a result of integration
of the anomaly of $U(1)$ R-symmetry (see \cite{Seiberg88} and \cite{DAdda}, correspondingly).
Surprisingly, the potential (\ref{3}) nicely correlates with the low-energy effective action of
4d $\cN=4$ SYM effective action in the $\cN=2$ superspace \cite{BuIv}. Indeed, the first term in the right-hand side
of (\ref{3}) formally coincides with the non-holomorphic potential ${\cal H}({\cal W},\bar{\cal
W})$ while the last term in (\ref{3}) is very similar to the
hypermultiplet completion of the non-holomorphic potential that was constructed in
\cite{BuIv}.

This analogy is even more striking. Indeed, in
\cite{Int} it was demonstrated that the low-energy effective
action in $\cN=4$ SYM theory contains the Wess-Zumino term for scalar fields which
originates from the 't~Hooft anomaly matching for the R-symmetry. This
Wess-Zumino term implies the non-renormalization of the coefficient
in front of the non-holomorphic potential beyond one loop. As we
show in this paper, the potential (\ref{3}) is also responsible
for the Wess-Zumino term for two-dimensional scalars, and exactly
the same arguments provide its non-renormalization.

One of the results of this paper is the illustration of the
deep interplay between the two-dimensional $\cN=(4,4)$ SQED and 4d
$\cN=4$ SYM theory at low energies, although they are very
different in general.

This paper is organized as follows. We start the main part of the
text (Section 2) with a short review of the gauge theory in
$\cN=(2,2)$ superspace and consider basic properties of the
parallel displacement propagator which is a key ingredient of the
technique of gauge-covariant perturbative computations
(for 4d gauge theories in $\cN=1$ superspace
this technique was developed in \cite{ILB82,Ohrndorf,K-Mc}
and for field theory on the supergravity background in \cite{McArthur83,McArthur84,BK1,BK2}).
Making use of this propagator, we construct exact Green's functions for chiral superfields
on covariantly constant vector multiplet background. In Section 3, we compute
the low-energy effective action in $\cN=(2,2)$ SQED with different
numbers of chiral multiplets. We start with a review of
old results \cite{DAdda} of one-loop quantum contributions to
the effective action and show how they can be naturally
reproduced by taking advantage of the technique of covariant perturbative computations
in the $\cN=(2,2)$ superspace. This technique is then applied to
compute two-loop quantum corrections to the effective action of
$\cN=(2,2)$ SQED in the Coulomb branch. In Section 4 we study the
structure of low-energy effective action in $\cN=(4,4)$ SQED to
the two-loop order in perturbation theory and discuss its
interplay with the 4d $\cN=2$ and $\cN=4$ SYM effective actions.
The Conclusions section is devoted to discussions of possible
extensions of the results of this work. In Appendix we give a
summary of our superspace conventions.

\section[Exact propagators on constant vector multiplet
background]{Exact propagators on constant vector multiplet \protect\\ background}
\setcounter{equation}0

In this section we consider two-dimensional non-Abelian gauge theory in
$\cN=(2,2)$ superspace and, following \cite{K-Mc}, we introduce
parallel displacement propagator which is a key ingredient of
gauge-covariant technique of multi-loop quantum computations.
Using the properties of this propagator we construct exact heat
kernels for basic Green's functions on covariantly constant vector multiplet
background. In the Abelian case, we apply these heat kernels in the
subsequent sections to compute two-loop quantum correction to the
effective action. We hope that the results of this section will be of
use also for the study of effective action in non-Abelian gauge
theories which will be considered elsewhere.
We keep the structure of this section close
to the corresponding presentation in \cite{3d4} to facilitate the
comparison with the three-dimensional gauge theory in $\cN=2$
superspace.

\subsection{Gauge theory in $\cN=(2,2)$ superspace}
\label{Gauge-sect}
We consider the two-dimensional $\cN=(2,2)$ superspace with
coordinates $z^A = (x^m,\theta^\alpha,\bar\theta_\alpha)$, where
$x^m$, $m=0,1$, are the Minkowski space coordinates, $\theta^\alpha$,
$\alpha=1,2$,
are Grassmann coordinates and $\bar\theta_\alpha =
(\theta_\alpha)^*$ are their conjugate. Our superspace conventions
are summarized in Appendix. They are chosen to be close to the
ones employed in the series of papers \cite{3d4,3d1,3d2,3d3,3d5}
devoted to the study of superfield theories in
three-dimensional $\cN=2$ superspace.

The (non-Abelian) gauge theory in the $\cN=(2,2)$ superspace is
described by the set of gauge-covariant superspace derivatives
\be
\nabla_A=
 (\nabla_m, \nabla_\alpha,\bar\nabla^\alpha) = D_A + V_A\,,
\label{V}
\ee
where $D_A =(\partial_m, D_\alpha,\bar D^\alpha)$ are
super-covariant derivatives, see (\ref{A3}), and $V_A=(V_m,V_\alpha,\bar V^\alpha)$ are
gauge connections subject to the constraints
\begin{subequations}
\label{algebra}
\bea
\{\nabla_\alpha,\bar\nabla_\beta \}&=&-2i(\gamma^m)_{\alpha\beta}
\nabla_m +2i\varepsilon_{\alpha\beta}G
+2\gamma^3_{\alpha\beta} H
\,,
\label{alg1}
\\
{}[\nabla_\alpha,\nabla_m]&=&-(\gamma_m)_{\alpha\beta}\bar
W^\beta\,,\qquad
[\bar\nabla_\alpha,\nabla_m]=(\gamma_m)_{\alpha\beta}
W^\beta\,,\label{alg2}\\
{}[\nabla_m,\nabla_n]&=&iF_{mn}\,.
\eea
\end{subequations}
Here  $G$, $H$, $W_\alpha$ and $F_{mn}$ are superfield strengths with
the following conjugation properties
\be
G^\dag=G\,,\quad
H^\dag = H\,,
\quad
(W_\alpha)^\dag=\bar W_\alpha\,,\quad
(F_{mn})^\dag=F_{mn}\,.
\ee
In two dimensions, the antisymmetric tensor $F_{mn}$ has
only one independent component, $F_{mn} = \varepsilon_{mn}f$,
where for the antisymmetric $\varepsilon$-tensor $\varepsilon_{mn}=-\varepsilon_{nm}$
with vector indices we use the convention $\varepsilon_{01} =
-\varepsilon^{01}=1$.

The algebra of covariant derivatives (\ref{algebra}) implies a
number of Bianchi identities. In particular, the scalar
superfields $G$ and $H$ are covariantly linear,
\be
\nabla^2 G = \nabla^2 H =0\,,\qquad
\bar \nabla^2 G = \bar \nabla^2 H=0\,.
\label{cov-lin}
\ee
The spinor superfield strengths $W_\alpha$ and $\bar W_\alpha$ are
expressed via the scalar ones,
\be
\bar W_\alpha=\nabla_\alpha G
             = -i(\gamma^3)_\alpha^\beta \nabla_\beta H \,,\qquad
W_\alpha=\bar \nabla_\alpha G = i(\gamma^3)_\alpha^\beta \bar \nabla_\beta H\,.
\label{W-G}
\ee
As usual, they have the chirality properties
\be
\nabla_\alpha\bar W_\beta=0\,,\qquad
\bar \nabla_\alpha W_\beta=0
\label{cov-chiral}
\ee
and obey the `standard' Bianchi identity
\be
\nabla^\alpha W_\alpha=\bar \nabla^\alpha\bar W_\alpha\,.
\ee
In its turn, the tensor field strength $F_{mn}$ is expressed via
$W_\alpha$ and $\bar W_\alpha$,
\be
F_{mn}\equiv \varepsilon_{mn} f = -\frac14 \varepsilon_{mn} (\gamma^3)^{\alpha\beta}(\nabla_\alpha W_\beta - \bar \nabla_\alpha \bar
W_\beta)\,.
\label{F-W}
\ee
Another important relation appears by commuting (\ref{alg1}) with
the superfield $G$ and applying properties (\ref{W-G})
\be
\nabla_\alpha W_\beta + \bar\nabla_\alpha \bar W_\beta=
 -2i\gamma^m_{\alpha\beta} \nabla_m G
  + 2 \gamma^3_{\alpha\beta}[H,G]
  +\varepsilon_{\alpha\beta} \nabla^\gamma W_\gamma\,.
\label{W-useful-id}
\ee

The algebra of covariant derivatives (\ref{algebra}) is invariant
under the $\tau$-gauge transformations
\be
\nabla_A \rightarrow e^{i\tau(z)} \nabla_A e^{-i\tau(z)}\,,
\ee
with $\tau(z)$ being real gauge superfield parameter, $\tau^\dag =
\tau$.

The gauge connections $V_A$ may be expressed via a prepotential.
In this paper we will use the real superfield prepotential $V$
which is introduced in such a way that the gauge-covariant spinor
derivatives acquire the form (chiral representation)
\be
\nabla_\alpha=e^{-2V}D_\alpha e^{2V}\,,\qquad
\bar\nabla_\alpha=\bar D_\alpha\,.
\ee
In this case, as a consequence of (\ref{alg1}), the scalar superfield strengths are expressed via the
prepotential as
\be
\label{VGH}
G=\frac i4\bar D^\alpha(e^{-2V} D_\alpha e^{2V})\,,\qquad
H=-\frac14(\gamma^3)^{\alpha\beta}\bar D_\beta( e^{-2V} D_\alpha e^{2V})\,.
\ee
The expressions of the other superfield strengths in terms of $V$ can
be obtained using (\ref{W-G}) and (\ref{F-W}). Note that all these
superfield strengths transform covariantly under the $\Lambda$-gauge
transformation of the prepotential
\be
e^{2V}\to e^{i\bar\Lambda}e^{2V}e^{-i\Lambda}\,,
\ee
with a chiral $\Lambda$.

The superfield strengths $G$ and $H$ can be considered as the real
and imaginary parts of a complex superfield $\Sigma$ and its (Hermitian)
conjugate $\bar\Sigma$
\be
\Sigma = G + i H\,,\qquad \bar\Sigma = G - i H\,.
\label{Sigma-def}
\ee
From (\ref{W-G}) it is easy to deduce twisted chirality
properties of these superfields
\be
\nabla_+ \Sigma = \bar\nabla_- \Sigma = 0\,,\qquad
\bar\nabla_+\bar\Sigma = \nabla_- \bar\Sigma =0\,,
\label{Sigma-tw-chiral}
\ee
where $(\nabla_+,\nabla_-) \equiv (\nabla_1, \nabla_2)$. The
existence of such twisted chiral superfields is an important
feature of two-dimensional gauge theory in superspace as compared
with the higher-dimensional cases. These superfield strengths
play central role in superfield description of gauge theories in
the $\cN=(2,2)$ superspace.

\subsection{Parallel displacement propagator in $\cN=(2,2)$ superspace}

In superspace, the parallel displacement propagator was introduced
in the work \cite{K-Mc} as a key ingredient which provides correct
transformation properties of Green's functions and corresponding
heat kernels under gauge transformations. This
allowed the authors of \cite{K-Mc} to develop a gauge-covariant procedure of
perturbative computations of effective actions in supersymmetric gauge
theories. In particular, this technique appeared very fruitful in
the study of low-energy effective actions in different
four-dimensional gauge theories in $\cN=1$ and $\cN=2$ superspaces
\cite{K1,K3,K4,K5,K6,K7,K-Tyler}. For three-dimensional gauge
theories this method was generalized in \cite{3d4,3d5}. This section
is aimed at extending the basic concepts of the procedure of
covariant perturbative computations to two-dimensional gauge
theories in the $\cN=(2,2)$ superspace.

Let us consider a superfield $\Phi$ in some representation $R$ of
the gauge group, and its Hermitian conjugate $\bar\Phi$
transforming in the representation $\bar R$,
\be
\Phi(z)\to \Phi'(z) = e^{i\tau(z)}\Phi(z)\,,\qquad
\bar\Phi(z) \to \bar\Phi'(z) = \bar\Phi(z) e^{-i\tau(z)}\,,
\ee
where $\tau = \tau^\dag$ is Hermitian, but otherwise arbitrary
gauge superfield parameter.
Correspondingly, Green's function for these fields $G(z,z')=
i\langle \Phi(z) \bar \Phi(z')\rangle$ has the transformation
property
\be
G(z,z') \to e^{i\tau(z)} G(z,z') e^{-i\tau(z')}\,.
\ee

In a similar way, the parallel displacement propagator $I(z,z')$
is, by definition, a two-point superspace function which transforms under the
gauge group as
\be
I(z,z') \to e^{i\tau(z)} I(z,z') e^{-i\tau(z')}\,.
\label{I-gauge}
\ee
Moreover, it is required to obey the differential equation
\be
\zeta^A \nabla_A I(z,z')= \zeta^A (D_A + V_A) I(z,z') = 0\,,
\label{zetaI}
\ee
and the boundary condition
\be
I(z,z) = {\bf 1}\,.
\label{I-boundary}
\ee
The latter means that at coincident superspace points $I(z,z')$ reduces to the
identity operator in the gauge group. In Eq.\ (\ref{zetaI}),
$\zeta^A\equiv (\rho^m,\zeta^\alpha,\bar\zeta_\alpha)$
is the $\cN=(2,2)$ supersymmetric interval with the components
\be
\rho^{m}=(x-x')^{m}
-i(\theta-\theta')^{\alpha} \gamma^m_{\alpha\beta}
 \bar\theta'^{\beta}
+i\theta'^{\alpha}\gamma^m_{\alpha\beta}(\bar\theta-\bar\theta')^{\beta}\,,\quad
\zeta^\alpha=(\theta-\theta')^\alpha\,,\quad
\bar\zeta_\alpha=(\bar\theta-\bar\theta')_\alpha\,.
\label{super-interval}
\ee

It is possible to show that
the properties (\ref{I-gauge}) and (\ref{I-boundary}) imply the
important relation
\be
I(z,z')I(z',z)  = {\bf 1}\,.
\ee
Note also that the rule of Hermitian conjugation for $I(z,z')$
looks like
\be
\left( I(z,z') \right)^\dag = I(z',z)\,.
\ee

The algebra of covariant derivatives (\ref{algebra}) can be
represented in the condensed form
\be
[ \nabla_A , \nabla_B \} = {\bf T}_{AB}{}^C \nabla_C + i {\bf
F}_{AB}\,,
\ee
where ${\bf T}_{AB}{}^C$ is the supertorsion and ${\bf F}_{AB}$ is the field
strength for gauge superfield connection (\ref{V}). The
non-vanishing components of these tensors can be read off from
(\ref{algebra}). They appear in the following important relation
for the derivative of the parallel displacement propagator \cite{K-Mc}
\bea
\label{dec2}
\nabla_B I(z,z')&=&i \sum_{n=1}^\infty \frac{(-1)^n}{(n+1)!}\,\bigg[-\zeta^{A_n}
 \ldots\zeta^{A_1}\nabla_{A_1}\ldots\nabla_{A_{n-1}}{\bf F}_{A_n B}(z)\\
&& + \frac{(n-1)}{2}\zeta^{A_n}{\bf T}_{A_n B}{}^{C} \zeta^{A_{n-1}}
 \ldots\zeta^{A_1}\nabla_{A_1}\ldots\nabla_{A_{n-2}}{\bf F}_{A_{n-1} C}(z)
 \bigg]I(z,z')\,.\nn
\eea
This identity shows that any covariant derivative of the parallel
displacement propagator may be expressed in terms of the parallel
displacement propagator itself and covariant derivatives of the
superfield strength together with the torsion tensor. This identity appears
crucial in perturbative computations of low-energy effective
action which is a functional of these tensors.

In general, (\ref{dec2}) is an infinite series over
covariant derivatives of the field strength ${\bf F}_{AB}$. It is
natural to expect that for certain field configurations this series
terminates. In particular, it is possible to show that for the
covariantly constant vector multiplet background
\be
\nabla_m \Sigma = \nabla_m \bar\Sigma =0\,,\qquad
\nabla_m W_\alpha = \nabla_m \bar W_\alpha =0
\ee
the identity (\ref{dec2}) reduces to
\begin{subequations}
\bea
\nabla_\beta I(z,z') &=& \bigg[ -i \bar\zeta_\beta G +
(\gamma^3)_\beta ^\alpha \bar\zeta_\alpha H
+\frac16(\gamma^m)_{\beta\alpha}\left(
3\rho_m \bar W^\alpha
- \rho_m \bar\zeta_\gamma \bar\nabla^\gamma \bar W^\alpha
- \bar\zeta^\alpha \rho^n F_{nm} \right) \nn\\&&
+\frac i6 \bar\zeta^2 W_\beta +\frac i6 \bar\zeta_\beta
\zeta^\alpha \bar W_\alpha
-\frac i3 \bar\zeta^\alpha \zeta_\alpha \bar W_\beta
-\frac i2 \bar\zeta_\alpha \zeta^\gamma (\gamma^3)_\beta^\alpha
(\gamma^3)_\gamma^\delta \bar W_\delta
\nn\\&&
+\frac i6 \bar\zeta^2 \zeta^\alpha
 \bar\nabla_{(\alpha} \bar W_{\beta)}\bigg] I(z,z')\,,\label{derI1}\\
\bar\nabla^\beta I(z,z') &=& \bigg[
-i \zeta^\beta G + (\gamma^3)_\alpha^\beta \zeta^\alpha H
+\frac 16 (\gamma^m)^{\beta\alpha}
\left( 3 \rho_m W_\alpha - \rho_m \zeta^\gamma \nabla_\gamma W_\alpha
 - \zeta_\alpha \rho^n F_{nm} \right)
\nn\\&&
-\frac i6 \zeta^2 \bar W^\beta - \frac i6 \zeta^\beta
\bar\zeta^\alpha W_\alpha
+\frac i3 \zeta^\alpha \bar\zeta_\alpha W^\beta
-\frac i2 \zeta^\alpha \bar\zeta_\gamma
 (\gamma^3)_\alpha^\beta (\gamma^3)_\delta^\gamma W^\delta
\nn\\&&
-\frac i6 \zeta^2 \bar\zeta_\alpha
 \nabla^{(\alpha} W^{\beta)}
\bigg]I(z,z')\,,\label{derI2}\\
\nabla_m I(z,z') &=& \bigg[
\frac i2 \rho^n F_{nm} -\frac12 (\gamma_m)_{\alpha\beta}
\Big(
\zeta^\alpha \bar W^\beta + \bar\zeta^\alpha W^\beta
 - \frac13 \zeta^\alpha \bar\zeta_\gamma \bar\nabla^\gamma \bar
 W^\beta
 \nn\\&&
+\frac13 \bar\zeta^\alpha \zeta_\gamma \nabla^\gamma W^\beta
\Big)
\bigg]I(z,z')\,.
\eea
\end{subequations}
As we will show in the following subsections, these identities
appear very useful in computing heat kernels of
Green's functions of various operators in the $\cN=(2,2)$
superspace.

\subsection{Real superfield Green's function and its heat kernel}

The real superfield d'Alembertian is defined by either expression
\bea
\square_{\rm v}&=&
\frac1{16}\{\nabla^2 , \bar\nabla^2 \}
-\frac18 \nabla^\alpha \bar\nabla^2 \nabla_\alpha
+\frac i2 (\nabla^\alpha W_\alpha) + i W^\alpha\nabla_\alpha
\nn\\
&=&\frac1{16}\{\nabla^2 , \bar\nabla^2 \}
-\frac18\bar\nabla^\alpha \nabla^2 \bar\nabla_\alpha
-\frac i2(\bar\nabla^\alpha \bar W_\alpha)
-i\bar W^\alpha \bar\nabla_\alpha\,.
\eea
By making use of the algebra (\ref{algebra}), this operator may be
brought to the form
\be
\square_{\rm v} = \nabla^m \nabla_m + \frac12\{\Sigma,\bar\Sigma\}
+iW^\alpha \nabla_\alpha
-i\bar W^\alpha \bar\nabla_\alpha\,.
\ee
Green's function $G_{\rm v}(z,z')$ of this operator is defined as
a solution of the equation
\be
(\square_{\rm v} +m^2 ) G_{\rm
v}(z,z')=-\delta^{2|4}(z-z')\,,
\label{Gv-eq}
\ee
where $m$ is a mass parameter and $\delta^{2|4}(z-z')$ is the
full superspace delta function,
\be
\delta^{2|4}(z-z') = \delta^2(x-x')\delta^4(\theta-\theta')\,.
\label{delta-full}
\ee
Green's function can be represented as a proper-time integral of
the corresponding heat kernel $K_{\rm v}(z,z'|s)$
\be
G_{\rm v}(z,z')=-i\int_0^\infty ds\, K_{\rm v}(z,z'|s)e^{-s(\epsilon+im^2)}\,,
\ee
where $\epsilon\to+0$ implements standard boundary condition for
the propagator. The equation for the propagator (\ref{Gv-eq}) is
satisfied when the heat kernel obeys the conditions
\be
(i\frac d{ds}-\square_{\rm v})K_{\rm v}(z,z'|s)=0\,,\qquad
\lim_{s\to0} K_{\rm v}(z,z'|s)=\delta^{2|4}(z-z')\,.
\ee
In general, it is very hard to solve these equations explicitly.
Nevertheless, it is possible to find the exact solution for the
heat kernel when the background gauge superfield obeys the
following two constraints:
\begin{itemize}
\item[i)] Gauge multiplet obeys super Yang-Mills equations of
motion (on-shell background)
\be
\nabla^\alpha W_\alpha=0\,,\qquad
\bar\nabla^\alpha \bar W_\alpha =0 \,;
\label{W-on-shell}
\ee
\item[ii)] Field strengths are covariantly constant
\be
\nabla_m \Sigma = \nabla_m \bar\Sigma =0\,,\qquad
\nabla_m W_\alpha = \nabla_m \bar W_\alpha =0\,.
\label{cov-const}
\ee
\end{itemize}
It is important to note that the compatibility condition for the constraint (\ref{cov-const})
requires that the background gauge superfield belongs to the Cartan
subalgebra of the Lie algebra of the gauge group. This means that
the background gauge superfields are (anti)commuting.

The procedure of solving the heat kernel equation for the
covariantly constant vector multiplet background was developed in the
four-dimensional case in \cite{McArthur83,McArthur84,Ohrndorf,K-Mc} and successfully applied for
three-dimensional gauge theories in \cite{3d4,3d1,3d2}. In the
two-dimensional case the same procedure yields
\be
K_{\rm v}(z,z'|s)=\frac1{4\pi s}
\frac{sf}{
\sinh(sf)}
e^{-is\Sigma\bar\Sigma}
{\cal O}(s)
e^{-\frac i4(f\coth s f)\rho^m\rho_m}
\zeta^2\bar\zeta^2 I(z,z')\,,
\label{Kv}
\ee
where $\zeta^2= \zeta^\alpha\zeta_\alpha$, $\bar\zeta^2 = \bar\zeta^\alpha
\bar\zeta_\alpha$ and ${\cal O}(s)$ is the `shift' operator
\be
{\cal O}(s)=e^{s(W^\alpha\nabla_\alpha
-\bar W^\alpha\bar\nabla_\alpha)}\,.
\ee

Within quantum loop computations, it is often necessary to know
the value of the
heat kernel at coincident superspace points. For this aim, it is
useful to have such a representation for the heat kernel (\ref{Kv}) where the
operator ${\cal O}(s)$ appears on the right and hits the parallel
displacement propagator,
\be
K_{\rm v}(z,z'|s)=\frac1{4\pi s}
\frac{s f}{
\sinh(s f)}
e^{-is\Sigma\bar\Sigma}
e^{-\frac i4(f\coth s f)\zeta^m(s)\zeta_m(s)}
\zeta^2(s)\bar\zeta^2(s) I(z,z'|s)\,.
\label{Kv2}
\ee
Here the operator ${\cal O}(s)$ is used to define the $s$-dependent
superfield strengths and components of the superspace interval
\begin{subequations}
\label{id's}
\bea
W^\alpha(s)&\equiv&{\cal O}(s)W^\alpha {\cal O}(-s)
=W^\beta (e^{sN})_\beta{}^\alpha\,,\\
\zeta^\alpha(s)&\equiv&{\cal O}(s)\zeta^\alpha {\cal O}(-s)
= \zeta^\alpha+W^\beta
((e^{sN}-1)N^{-1})_\beta{}^\alpha\,,\\
\bar\zeta^\alpha(s)&\equiv&{\cal O}(s)\bar\zeta^\alpha {\cal O}(-s)
= \bar\zeta^\alpha-\bar W^\beta
((e^{-s\bar N}-1)\bar N^{-1})_\beta{}^\alpha\,,\\
\rho^m(s)&\equiv&{\cal O}(s)\rho^m{\cal O}(-s)
=\rho^m+i(\gamma^m)^{\alpha\beta}\int_0^s dt\left(W_{\alpha}(t)\bar\zeta_{\beta}(t)
+\bar W_{\alpha}(t)\zeta_{\beta}(t)\right)\,,~~~~~~~~
\eea
\end{subequations}
and
\be
I(z,z'|s)\equiv {\cal O}(s)I(z,z')\,.
\label{Is}
\ee
In formulas (\ref{id's}) we have introduced the notation
\be
N_{\alpha\beta} = \nabla_{(\alpha} W_{\beta)}\,,\qquad
\bar N_{\alpha\beta} = \bar\nabla_{(\alpha} \bar W_{\beta)}\,.
\ee

The $s$-dependent parallel displacement propagator
(\ref{Is}) can be represented in the following form \cite{K-Mc}
\be
I(z,z'|s) = \exp\left[
\int_0^s dt\, \Xi(z,z'|t)
\right]I(z,z')\,,
\label{Is2}
\ee
where
\be
\Xi(z,z'|t) = {\cal O}(t) \Xi(z,z'){\cal O}(-t)\,,
\ee
and $\Xi(z,z')$ solves for
\be
(W^\alpha\nabla_\alpha
-\bar W^\alpha\bar\nabla_\alpha)I(z,z') = \Xi(z,z')I(z,z')\,.
\ee
Making use of (\ref{derI1}) and (\ref{derI2}) we find
\bea
\Xi(z,z')
&=&i(\zeta^\alpha \bar W_\beta + \bar\zeta_\beta W^\alpha)
(\delta_\alpha^\beta G + i(\gamma^3)_\alpha^\beta H)
%
\nn\\
&&+\frac i6 \bar\zeta^2 ( W^2 + \zeta^\alpha W^\beta \nabla_\alpha
W_\beta)
+\frac i6 \zeta^2 ( \bar W^2 + \bar\zeta^\alpha \bar W^\beta \nabla_\alpha
W_\beta)
\nn\\&&
+\frac{4 i}3\zeta^\alpha \bar\zeta^\beta W_\alpha \bar W_\beta
- i \zeta^\alpha \bar\zeta_\alpha W^\beta\bar W_\beta
+i \zeta^\gamma \bar\zeta_\delta
 (\gamma^3)^\alpha_\beta (\gamma^3)_\gamma^\delta
  W^\beta \bar W_\alpha\,.
\label{Xi}
\eea
The expression for $\Xi(z,z'|s)$ can be found from the above
formula just by replacing all superfield strengths and components
of the superspace interval by the corresponding $s$-dependent
quantities from (\ref{id's}).

\subsection{Heat kernel for chiral superfield Green's function}

Consider gauge-covariant chiral superfield $\Phi$,
$\bar\nabla_\alpha\Phi=0$, and its Hermitian conjugate $\bar\Phi$. The
d'Alembertian operators acting in the space of such fields are
defined in the standard way
\be
\square_+ \Phi = \frac1{16} \bar\nabla^2 \nabla^2 \Phi\,,\qquad
\square_- \bar\Phi = \frac1{16} \nabla^2 \bar\nabla^2\bar\Phi\,.
\label{box+-def}
\ee
Making use of the algebra of covariant derivatives (\ref{algebra})
one uncovers the following representations for these
operators
\begin{subequations}
\label{box+-}
\bea
\square_+&=&
\frac1{16} \bar\nabla^2 \nabla^2=
\nabla^m\nabla_m+\frac12\{\Sigma,\bar\Sigma\}+\frac i2(\nabla^\alpha W_\alpha)
+iW^\alpha\nabla_\alpha\,,
\label{box+}\\
\square_-&=&
\frac1{16}\nabla^2 \bar\nabla^2
=\nabla^m\nabla_m+\frac12\{\Sigma,\bar\Sigma\}-\frac i2(\bar\nabla^\alpha \bar W_\alpha)
-i\bar W^\alpha\bar\nabla_\alpha\,.
\label{box-}
\eea
\end{subequations}
By definition, Green's functions for these operators and the corresponding propagators
obey
\begin{subequations}
\label{Green'sFunctions}
\begin{align}
i\langle \Phi(z)\Phi^{\rm T}(z') \rangle &= -mG_+(z,z')\,, &
(\square_+ +m^2) G_+(z,z')&=-\delta^{2|2}_+(z,z')\,,\label{G+def}\\
i\langle \bar\Phi^{\rm T}(z)\bar \Phi(z') \rangle &= mG_-(z,z')\,,&
(\square_- +m^2) G_-(z,z')&=-\delta^{2|2}_-(z,z')\,,
\label{Gpm-eqs}
\end{align}
\end{subequations}
where $\delta^{2|2}_\pm(z,z')$ are (anti)chiral delta-functions
which are related to the full superspace delta-function (\ref{delta-full}) as
\be
\delta_+^{2|2}(z,z') = -\frac14 \bar\nabla^2
\delta^{2|4}(z-z')\,,\qquad
\delta_-^{2|2}(z,z') = -\frac14 \nabla^2 \delta^{2|4}(z-z')\,.
\ee
For Green's functions (\ref{Green'sFunctions}) there are the associated heat
kernels
\be
G_\pm(z,z')=-i\int_0^\infty ds\, K_\pm(z,z'|s)e^{
-s(\epsilon+im^2)}\,,\qquad
\epsilon\to+0\,.
\label{K+-def}
\ee

It is known \cite{K-Mc,K1} that for the on-shell vector multiplet
background (\ref{W-on-shell}) the chiral Green functions $G_\pm$ are related to
$G_{\rm v}$ as
\be
G_+(z,z')=-\frac14\bar \nabla^2 G_{\rm v}(z,z')\,,\qquad
G_-(z,z')=-\frac14\nabla^2 G_{\rm v}(z,z')\,.
\label{G-relations}
\ee
It is easy to check these relations using the identities
\bea
\nabla^2\square_+=\square_-\nabla^2\,,&\qquad&
\bar\nabla^2\square_-=\square_+\bar\nabla^2\,,\\
\nabla^2\square_+=\nabla^2\square_{\rm v}=\square_{\rm
v}\nabla^2\,,&\qquad&
\bar\nabla^2\square_-=\bar\nabla^2\square_{\rm v}=\square_{\rm v}\bar\nabla^2\,.
\label{last}
\eea
It should be noted that the identities (\ref{last}) hold only for the
on-shell vector multiplet background (\ref{W-on-shell}). The
equations (\ref{G-relations}) imply similar relations for the
corresponding heat kernels
\be
K_+(z,z'|s)=-\frac14 \bar \nabla^2 K_{\rm v}(z,z'|s)\,,\qquad
K_-(z,z'|s)=-\frac14 \nabla^2 K_{\rm v}(z,z'|s)\,.
\ee
Thus, the computation of the heat kernels $K_\pm$ is reduced to finding
the result of the action of the operators $\nabla^2$ and $\bar\nabla^2$ on the heat
kernel (\ref{Kv}).

It is possible to show that upon acting by $\bar\nabla^2$ on
(\ref{Kv}), this operator hits only $\zeta^2\bar\zeta^2
I(z,z')$ since the factor in front of this function originates from $e^{-is \square_{\rm v}}$. The latter
operator commutes with $\bar\nabla^2$ owing to the identities
(\ref{last}). Thus, for $K_+$ we have
\be
K_{+}(z,z'|s)=\frac1{4\pi s}
\frac{sf}{
\sinh(sf)}
e^{-is\Sigma\bar\Sigma}
{\cal O}(s)
e^{-\frac i4(f\coth s f)\rho^m\rho_m}
\zeta^2 \left(-\frac14 \bar\nabla^2\right) \bar\zeta^2 I(z,z')\,.
\label{K+1}
\ee
Applying (\ref{derI2}) we compute the action of the operator
$\bar\nabla^2$ on the parallel displacement propagator
\be
-\frac14 \zeta^2 \bar\nabla^2(\bar\zeta^2 I(z,z'))=\zeta^2
e^{-\frac12(\gamma^m)_{\alpha\beta}\rho_m \bar\zeta^\alpha
W^\beta} I(z,z')\,.
\ee
Substituting this identity into (\ref{K+1}) we find
\bea
K_{+}(z,z'|s)&=&\frac1{4\pi s}
\frac{sf}{
\sinh(sf)}
e^{-is\Sigma\bar\Sigma}
e^{-\frac i4(f\coth s f)\rho^m(s)\rho_m(s)
-\frac12(\gamma^m)_{\alpha\beta}\rho_m(s) \bar\zeta^\alpha(s)
W^\beta(s)}
\nn\\&&\times
\zeta^2(s)  I(z,z'|s)\,.
\label{K+}
\eea
Here we pushed the operator ${\cal O}(s)$ through on the right
that resulted in making all objects $s$-dependent according to
(\ref{id's}) and (\ref{Is}).
In a similar way we find the antichiral heat
kernel
\bea
K_{-}(z,z'|s)&=&\frac1{4\pi s}
\frac{sf}{
\sinh(sf)}
e^{-is\Sigma\bar\Sigma}
e^{-\frac i4(f\coth s f)\rho^m(s)\rho_m(s)
-\frac12(\gamma^m)_{\alpha\beta}\rho_m(s) \zeta^\alpha(s)
\bar W^\beta(s)}
\nn\\&&\times
\bar\zeta^2(s)  I(z,z'|s)\,.
\label{K-}
\eea
We point out that the expressions for the (anti)chiral heat
kernels are very similar to the ones in the four-dimensional
supersymmetric gauge theory \cite{K1}.

\subsection{Heat kernel for Green's function $G_{+-}$}

Finally, we consider the propagators among chiral and antichiral
superfields
\be
i\langle \Phi(z) \bar\Phi(z') \rangle = G_{+-}(z,z')\,,\qquad
i\langle \bar\Phi(z) \Phi(z') \rangle = G_{-+}(z,z')\,.
\label{G+-def}
\ee
By definition, these Green's functions obey
\begin{subequations}
\bea
\frac14 \nabla^2 G_{+-}(z,z') + m^2 G_-(z,z') =
-\delta^{2|2}_-(z,z')\,,
\label{G+-def-a}
\\
\frac14\bar \nabla^2 G_{-+}(z,z') + m^2 G_+(z,z') =
-\delta^{2|2}_+(z,z')\,.
\eea
\end{subequations}
With Green's functions (\ref{G+-def}) are associated the corresponding heat
kernels
\begin{subequations}
\label{K2+-def}
\bea
\label{K+-def-a}
G_{+-}(z,z') &=& -i \int_0^\infty ds\, K_{+-}(z,z'|s)
e^{-s(\epsilon+im^2)}\,,
\\
G_{-+}(z,z') &=& -i \int_0^\infty ds\, K_{-+}(z,z'|s)
e^{-s(\epsilon+im^2)}\,.
\eea
\end{subequations}
This subsection aims to find explicit solutions for these heat
kernels on the covariantly constant vector multiplet background.

First of all, we point out that, as a consequence of the
definitions of covariantly (anti)chiral d'Alembertian operators
(\ref{box+-}), Green's functions (\ref{G+-def}) are related to the
(anti)chiral ones (\ref{Green'sFunctions}) as
\be
G_{+-}(z,z') = \frac14\bar\nabla^2 G_-(z,z')\,,\qquad
G_{-+}(z,z') = \frac14 \nabla^2 G_+(z,z')\,.
\ee
Analogous relations hold for the corresponding heat kernels
\be
K_{+-}(z,z'|s) = \frac14\bar\nabla^2 K_-(z,z'|s)\,,\qquad
K_{-+}(z,z'|s) = \frac14 \nabla^2 K_+(z,z'|s)\,.
\ee
Thus, the problem is reduced to finding the action of the
operators $\nabla^2$ and $\bar\nabla^2$ on the heat kernels
(\ref{K+}) and (\ref{K-}).

Let us consider the derivation of the heat kernel $K_{+-}$ in some
details. It appears upon acting by the operator $\bar\nabla^2$
on the heat kernel (\ref{K-}). Note that, owing to (\ref{last}),
this operator commutes with the expression $e^{-is\Sigma\bar\Sigma} e^{-\frac i4(f\coth s
f)\rho^m\rho_m}$ since the latter originates from $e^{-is \square_{\rm v}}$.
Thus, we need to find the action of this operator on the rest using
the properties of the parallel displacement operator (\ref{derI2})
\be
-\frac14 \bar\nabla^2 \left(
e^{-\frac12(\gamma^m)_{\alpha\beta}\rho_m \zeta^\alpha
\bar W^\beta}
\bar\zeta^2  I(z,z') \right)
= e^{R(z,z')}I(z,z')\,,
\ee
where
\bea
R(z,z')
&=& -i \zeta^\alpha \bar\zeta_\alpha G + \zeta^\alpha
\bar\zeta_\beta (\gamma^3)_\alpha^\beta H
-\frac12 (\gamma_m)_{\alpha\beta} \tilde\rho^m (\zeta^\alpha \bar W^\beta + \bar\zeta^\alpha W^\beta)
+\frac{2 i}3 \bar\zeta^2 \zeta^\alpha W_\alpha
\nn\\&&
-\frac i6 \zeta^2 \bar\zeta^\alpha \bar W_\alpha
+\frac12 (\gamma_m)^{\alpha\beta} \tilde\rho^m \zeta_\beta
\bar\zeta^\gamma \nabla_\alpha W_\gamma\,.
\label{R}
\eea
Here
\be
\tilde\rho^m = \rho^m + i \zeta^\alpha \gamma^m_{\alpha\beta}
\bar\zeta^\beta
\label{tilde-rho}
\ee
is a modification of the supersymmetric interval which is chiral
with respect to the first argument and antichiral with respect to
the other
\be
D'_\alpha \tilde\rho^m = \bar D_\alpha \tilde\rho^m =0\,.
\ee
Given the function $R(z,z')$ in the form (\ref{R}), we have the
following representation for the heat kernel $K_{+-}$
\be
K_{+-}(z,z'|s)=-\frac1{4\pi s}
\frac{sf}{
\sinh(sf)}
e^{-is\Sigma\bar\Sigma}
{\cal O}(s)
e^{-\frac i4(f\coth s f)\tilde\rho^m \tilde\rho_{m}
+R(z,z')}
I(z,z')\,.
\label{K+-}
\ee

As the final step, in (\ref{K+-}) we have to push the operator ${\cal
O}(s)$ through on the right and hit the parallel displacement
propagator according to Eq.\ (\ref{Is}). This
procedure effectively makes the superfield strengths and components
of supersymmetric interval $s$-dependent according to (\ref{id's})
\be
K_{+-}(z,z'|s)=-\frac1{4\pi s}
\frac{sf}{
\sinh(sf)}
e^{-is\Sigma\bar\Sigma}
e^{-\frac i4(f\coth s f)\tilde\rho^m(s)\tilde\rho_{m}(s)
+R(z,z'|s)+\int_0^s dt\, \Xi(t)}
  I(z,z')\,,
\label{K+-1}
\ee
where $R(z,z'|s) = {\cal O}(s)R(z,z'){\cal O}(-s)$, and $\Xi(s)$
is given by (\ref{Xi}). For practical computations, it is useful
to represent the heat kernel (\ref{K+-1}) in the equivalent form
\be
K_{+-}(z,z'|s)=-\frac1{4\pi s}
\frac{sf}{
\sinh(sf)}
e^{-is\Sigma\bar\Sigma}
e^{-\frac i4(f\coth s f)\tilde\rho^m(s)\tilde\rho_{m}(s)
+R(z,z')+\int_0^s dt(R'(t)+ \Xi(t))}
  I(z,z')\,.
\label{K+-2}
\ee
Here, the function $R'(t)$ can be found explicitly from (\ref{R})
using $R'(t)={\cal O}(t)[W^\alpha\nabla_\alpha
-\bar W^\alpha\bar\nabla_\alpha,R]{\cal O}(-t)$ and combined with
(\ref{Xi}):
\bea
R'(t)+\Xi(t)&=&{\cal O}(t)
\big[
2i \bar\zeta_\alpha W^\alpha G
- 2 (\gamma^3)_\alpha^\beta \bar\zeta_\beta W^\alpha H
+\frac{5i}{3} \zeta^\alpha W_\alpha \bar \zeta^\beta \bar W_\beta
\nn\\&&
+2i\zeta^\alpha \bar W_\alpha \bar \zeta^\beta W_\beta
-\frac i6\bar\zeta^2 W^2
+2i(\gamma^3)_\alpha^\beta (\gamma^3)_\gamma^\delta
 \zeta^\gamma \bar\zeta_\delta W^\alpha \bar W_\beta
\nn\\&&
-\frac{11i}{12}\bar\zeta^2 \zeta^\beta \nabla_\beta W^2
+\frac12 (\gamma_m)^{\alpha\beta}\tilde\rho^m
 \bar\zeta_\alpha \nabla_\beta W^2
\big]
{\cal O}(-t)\,.
\label{R+Xi}
\eea
We point out that, as follows from (\ref{R}), the function $R(z,z')$
vanishes at coincident Grassmann coordinates,
$R(z,z')|_{\zeta\to0}\to0$. However, the contribution from
(\ref{R+Xi}) is non-trivial at coincident points.

\section{Low-energy effective action in $\cN=(2,2)$ SQED}
\setcounter{equation}0

\subsection{General remarks}
In general, Abelian gauge theories in $\cN=(2,2)$ superspace may
include the following terms in the classical action:
\begin{itemize}
\item The kinetic term for the vector multiplet $V$
\be
S_V=\frac1{2e^2} \int d^{2|4}z \,\bar\Sigma\Sigma\,.
\label{Sv}
\ee
Here $e$ is the dimensional gauge coupling, $[e]=1$, and $d^{2|4}z$ is the measure in the
full $\cN=(2,2)$ superspace (see Appendix for our superspace
conventions).

\item The mass term for the vector multiplet
\be
S_{\mathfrak{m}}
= -\frac i4\frac{\mathfrak{m}}{e^2} \int d^{2|2} \tilde z\, \Sigma^2
+ c.c.\,,
\label{Svm}
\ee
where the integration goes over the twisted chiral subspace and ${\mathfrak m}$ is, in general, complex mass parameter.
Without loss of generality, we can set it to be real,
$\bar{\mathfrak m} = {\mathfrak m}$, just to simplify some formulas below.
Note that the sum of actions (\ref{Sv}) and (\ref{Svm}) amounts to the
massive Wess-Zumino model for the twisted chiral multiplet $\Sigma$.
It should be noted that the mass term
(\ref{Svm}) may be obtained by the dimensional reduction from the
3d $\cN=2$ Chern-Simons action which plays role of the
topological mass term in three-dimensional electrodynamics.

\item Fayet-Iliopoulos (FI) term
\be
S_{\rm FI} = -\frac{it}2 \int d^{2|2}\tilde
z\,\Sigma
+c.c.\,,
\label{SFI2}
\ee
where
\be
t= r+i\theta\,.
\label{t-complex}
\ee
In (\ref{SFI2}), the real part of the FI parameter $r$
couples with the auxiliary field $D$ of the vector multiplet while the imaginary part $\theta$
corresponds to the topological $f$-term and
quantizes \cite{Witten93,Coleman}.

\item $N$ charged chiral multiplets $Q_i$ with charges $q_i$ and
mass matrix $m_{ij}$
\be
S_Q = - \sum_{i=1}^N \int d^{2|4} z\, \bar Q_i e^{2q_i V} Q_i
-\sum_{i,j=1}^N \left (\int d^{2|2} z \, m_{ij} Q_i Q_j + c.c.\right)
\,.
\label{SQi}
\ee
Needless to say that the mass matrix $m_{ij}$ should be such that
the gauge invariance is preserved.
Chiral multiplets may also have real mass which can be absorbed by
shifts of scalars in the vector multiplet $V$.
\end{itemize}

More generally, it is also possible to study quantum dynamics of twisted chiral
multiplets as well as semi-chiral ones \cite{GHR,semichiral1,semichiral2,semichiral3},
but such models are beyond the scope of this paper.

Depending on the number of chiral multiplets and on the values of
all mentioned above parameters, Abelian gauge theories in the
$\cN=(2,2)$ superspace exhibit different phases which are
thoroughly investigated in \cite{Witten93}. In this paper, we are
interested in the effective action in the Coulomb branch. It is
known that the necessary condition for existence of the Coulomb
branch at the quantum level is that the charges of all chiral
multiplets should sum to zero
\be
\sum_{i=1}^N q_i =0\,.
\label{charge-zero}
\ee
Indeed, when this condition is not satisfied, the following two
effects occur: i) There are UV-divergent tadpole Feynman graphs which
result in the renormalization of the FI parameter. These quantum
corrections lift the Coulomb branch. ii) The
effective twisted superpotential for the superfield strength
$\Sigma$ is generated at one loop \cite{DAdda}. This effective superpotential
may also be interpreted as a functional reproducing correct
transformation properties of the effective action under anomalous
R-symmetry. Correspondingly, when the condition (\ref{charge-zero}) is
satisfied, there are no divergent quantum contributions to the FI
parameter and classical Coulomb branch is preserved at the quantum
level. The latter case is of primary importance for our studies as
we are interested in the two-loop quantum contributions to the
effective action in the Coulomb branch. However, in this section,
for the sake of completeness we will shortly consider a model for which the condition
(\ref{charge-zero}) is not satisfied and will give a superfield
derivation of the effective twisted superpotential obtained
originally in \cite{DAdda} by component field quantum
computations.

The typical example of the models for which the constraint
(\ref{charge-zero}) is violated is the supersymmetric
electrodynamics with one chiral multiplet while the well-known case
when this constraint is satisfied is the supersymmetric
electrodynamics with two chiral multiplets carrying opposite
charges under the $U(1)$ gauge symmetry. The latter will be
studied in Section \ref{two-fl-sec} while the former is considered just below.

\subsection{SQED with one chiral flavor}
\label{sec-SQED1}
\subsubsection{Classical action}
In this section, we consider the supersymmetric electrodynamics
with one chiral multiplet carrying charge $+1$
\be
\label{S1}
S = \int d^{2|4}z \left(
 \frac1{2e^2}\Sigma\bar\Sigma  - \bar Q
e^{2 V} Q\right)
-\left[ \frac i2 \int d^{2|2}\tilde z \left( t
\Sigma + \frac{{\mathfrak m}}{2e^2}\Sigma^2\right)
+c.c.\right].
\ee
Obviously, in the limit $e\to\infty$ the classical action
becomes scale invariant and superconformal, though this symmetry
is known to be broken by quantum corrections \cite{DAdda}.

Recall that the $\cN=(2,2)$ vector multiplet
contains a complex scalar $\sigma$ associated with the lowest
component of the superfield $\Sigma$
\be
\sigma \equiv \Sigma|\,,
\ee
where the bar-projection means vanishing $\theta$-variables.
Denoting the scalar fields in the chiral multiplet by
\be
\varphi \equiv Q |\,,\qquad
\bar\varphi \equiv \bar Q|\,,
\ee
it is not hard to find the scalar potential which appears after
elimination of auxiliary fields
\be
\mathscr{V} = \frac{e^2}{2} (\varphi\bar\varphi - t
 - \frac{\mathfrak m}{e^2} {\rm Re}\,\sigma
)^2
+ \varphi\bar\varphi \sigma\bar\sigma\,.
\label{V-pot}
\ee

The Coulomb branch is parametrized by the vev of the scalar field $\sigma$ in
the vector multiplet while the scalars from the chiral multiplet must
have vanishing vevs
\be
\mbox{Coulomb branch:}\qquad
\langle \sigma \rangle =const,\quad
\langle \varphi \rangle =0\,.
\label{ColBranch}
\ee
The vanishing of the scalar potential (\ref{V-pot})
for such values of scalars is possible only for special value of the FI parameter
\be
t =
- \frac{\mathfrak m}{e^2} \langle {\rm Re}\,\sigma \rangle
\,.
\label{t-zero}
\ee
In this case the chiral multiplet acquires real mass proportional to
the of vev of $\sigma$ while the vector multiplet (`photon')
has a small mass $\mathfrak m$. Naively, one could study the effective
action for the vector multiplet which appears by
integrating out the massive chiral multiplet. However, the constraint
(\ref{t-zero}) appears to be ruined by one-loop quantum corrections and the
Coulomb branch is lifted at the quantum level \cite{Witten93}.
Although this scenario is well-known, we will demonstrate it
explicitly by computing one-loop effective action in the model
(\ref{S1}). The details of these computations will be of use
in subsequent sections.

In general, the effective action for the vector multiplet $V$ may
have odd and even parts with respect to the reflection $V\to -V$,
\be
\Gamma[V] = \Gamma_{\rm odd}[V] + \Gamma_{\rm even}[V]\,,
\ee
where
\be
\Gamma_{\rm odd}[-V] = -\Gamma_{\rm odd}[V]\,,\qquad
\Gamma_{\rm even}[-V] = \Gamma_{\rm even}[V]\,.
\ee
Treatment of these parts in the effective action requires slightly
different computational methods. Therefore, we will consider them
separately.

\subsubsection{Even part of the one-loop effective action}
\label{even-eff-action}
Let $H$ be the operator which appears in the matrix of second variational derivatives of $S$
with respect to the chiral superfields,
\be
H=
\left(
\begin{array}{cc}
0 & \frac14\bar \nabla^2 \\
\frac14 \nabla^2  &0
\end{array}
\right).
\label{Hoperator}
\ee
The even part of the one-loop effective action $\Gamma$ may be
found by evaluating trace of logarithm of square of this operator
\be
\Gamma_{\rm even}=\frac i4 \Tr\ln H^2
=\frac i4\Tr \ln \square_+ +
c.c.
\label{G1}
\ee
Here we have taken into account the definition (\ref{box+-def}) of the chiral covariant
d'Alembertian in terms of covariant spinor derivatives.
Associated with this operator is the Green function $G_+(z,z')$ defined
in (\ref{G+def}) and the corresponding heat kernel $K_+(z,z'|s)$, see Eq.\
(\ref{K+-def}). Thus, for the effective action (\ref{G1}) we have
the following proper-time representation
\be
\Gamma_{\rm even}=-\frac i4\int_0^\infty\frac{ds}{s}{\rm Tr}_+\,K_+(s)
+c.c.\,,
\label{Geven}
\ee
where ${\rm Tr}_+ K_+(s)$ means the heat kernel $K_+(z,z'|s)$ at
coincident superspace points, $z'=z$, and integrated over the
chiral subspace
\be
{\rm Tr}_+\, K_+(s)=\int d^{2|2}z\,K_+(z,z|s)\,.
\ee
This reduces the problem of computation of the even part of the
effective action to evaluating the limit of coincident superspace
points for the heat kernel, $\lim_{z'\to z}K_+(z,z'|s)$.

Recall that we consider the low-energy effective action for the
on-shell, constant vector multiplet background specified by the
constraints (\ref{W-on-shell}) and (\ref{cov-const}). For this
background, the heat kernel $K_+$ was found in the form
(\ref{K+}). This formula involves different $s$-dependent objets
defined in (\ref{id's}) and (\ref{Is}). For the one-loop effective
action we need the values of these objets at coincident superspace
points when all components of the superspace interval vanish, $\zeta^A \to
0$. In particular, simple calculations yield
\be
\zeta^2(s)\Big|_{\zeta^A=0}=s^2 W^2\frac{\sinh^2\frac{s
f}{2}}{(sf/2)^2}\,,
\label{zeta0}
\ee
where $f$ is the component of the superfield strength tensor,
$F_{mn}= \varepsilon_{mn} f$, which can be regarded as
a constant for the considered background. It is important to note
that the formula (\ref{zeta0}) contains $W^2$ that prevents any
other contributions from the other $s$-depended objets in
(\ref{K+}). Thus, this kernel acquires simple form at coincident
superspace points
\be
K_+(z,z|s)
=\frac{1}{4\pi}s W^2 e^{-is\Sigma\bar\Sigma}
\frac{\tanh (sf/2)}{sf/2}\,.
\label{K+lim}
\ee
Substituting this expression into (\ref{Geven}) we find the
even part of the one-loop effective action
\be
\Gamma_{\rm even}=-\frac i{16\pi}\int d^{2|2}z
\int_0^\infty ds\, W^2e^{-is\Sigma\bar\Sigma}\frac{\tanh(sf/2)}{sf/2} + c.c\,.
\label{G-chiral}
\ee

It is instructive to rewrite the functional
(\ref{G-chiral}) in the full superspace
\bea
\Gamma_{\rm even} &=& \frac1{8\pi} \int d^{2|4}z\, \ln\Sigma \ln\bar\Sigma
\nn\\&&
+\frac i{8\pi} \int d^{2|4}z \int_0^\infty ds\, e^{-is\Sigma\bar\Sigma}
\frac{W^2 \bar W^2}{f^2}\left(
\frac{\tanh(sf/2)}{sf/2} -1
\right)\,.
\label{G-full}
\eea
The term in the first line here specifies the effective K\"ahler
potential for the twisted chiral superfield $\Sigma$. The term in the last
line in (\ref{G-full}) takes into account all higher-derivative
corrections with respect to the gauge superfield which can be
considered as the Euler-Heisenberg effective action.

We point out that the effective action (\ref{G-full}) was found
for the first time in \cite{DAdda} using component filed one-loop computations
and in \cite{Shizuya} by means of superfield methods.
Here we just gave a derivation of this effective action by taking advantage of the
superfield heat kernel technique. This result will be useful in the
study of low-energy effective action in the model with two chiral
flavors which will be considered in Section \ref{two-fl-sec}.

\subsubsection{Odd part of the one-loop effective action}
\label{odd-eff-action}
The odd part of the effective action cannot be found upon squaring
of the operator (\ref{Hoperator}). Instead, to catch up the odd
contributions we have to consider the general variation of the
effective action with respect to the vector multiplet
\be
\delta \Gamma = \int d^{2|4}z \, \delta V
\langle J \rangle  \,,
\label{VarGam}
\ee
where $\langle J \rangle$ is the effective current. In the
one-loop approximation, this effective
current receives contributions only from the chiral superfield
propagator in the vector multiplet background
\be
 \langle J \rangle  = -2 \langle  Q e^{2V} \bar Q \rangle
 =2iG_{+-}(z,z)\,,
\ee
where $G_{+-}(z,z')$ is defined in Eq.\ (\ref{G+-def-a}), with the mass parameter
set to zero, $m=0$. It is
useful also to represent this effective current via the heat
kernel $K_{+-}$ using (\ref{K+-def-a})
\be
 \langle J \rangle
 =2 \int_0^\infty ds\, K_{+-}(z,z|s)\,.
\label{J-K}
\ee
Thus, the computation of (\ref{VarGam}) is reduced to finding the trace of the heat
kernel $K_{+-}(z,z'|s)$.

The problem of evaluating the trace of the heat kernel
$K_{+-}(z,z'|s)$ is rather technically involved.
However, to find the odd
part of the effective action we don't actually need to now the
full expression for $K_{+-}(z,z|s)$. Indeed, the
full expression for $K_{+-}(z,z|s)$ contains different terms
which are responsible
both for odd and even parts of the effective action. Since the
even part of the effective action has been fully studied in the previous subsection, here we have
to focus only on possible contributions to $\Gamma_{\rm odd}$
from $K_{+-}(z,z|s)$. For this goal it is sufficient to
approximate $K_{+-}(z,z|s)$ by the terms with no derivatives of
 $\Sigma$,
\be
K_{+-}(z,z|s)\approx
-\frac1{4\pi s}
e^{-is\Sigma\bar\Sigma} \,.
\label{K-approx}
\ee
Substituting (\ref{K-approx}) into (\ref{J-K}) we have
UV-divergent integral over the proper time $s$. Introducing a
small regularization parameter $\epsilon$ this integral may be
evaluated
\be
\langle J\rangle =-\frac1{2\pi}\left(\frac1\epsilon - \gamma\right)
+ \frac 1{2\pi} \ln(\Sigma\bar\Sigma)\,,
\label{J1}
\ee
where $\gamma$ is the Euler-Mascheroni constant.
Thus, we see that the odd part of the effective action is the sum
of divergent and finite contributions
\be
\Gamma_{\rm odd} = \Gamma_{\rm div} + \Gamma_{\rm fin}\,.
\ee

The divergent part of the effective action can be immediately read
off from the first term in (\ref{J1})
\be
\Gamma_{\rm div} = -\frac1{2\pi \epsilon} \int d^{2|4}z \,
V
=\frac i{8\pi\epsilon}\int d^{2|2}\tilde z \,\Sigma +c.c.
\label{G-div}
\ee
This expression, being added to the classical action (\ref{S1}),
leads to infinite renormalization of the FI parameter
\be
t \to t' = t - \frac1{4\pi\epsilon}\,.
\label{t-inf}
\ee
This means that even if we switch off the FI parameter
classically, it is always generated by one-loop divergent tadpole
diagrams. This is the origin of lifting of the classical Coulomb
branch by quantum corrections advocated in \cite{Witten93}.

Substituting (\ref{J1}) into (\ref{VarGam}) we get the variation of finite terms in the odd part of
the effective action
\be
\delta\Gamma_{\rm fin} =\frac 1{2\pi} \int d^{2|4}z \,
\delta V \ln(\Sigma\bar\Sigma)\,.
\ee
Integrating this variation we uncover the effective twisted
potential for $\Sigma$
\be
\Gamma_{\rm fin} = -\frac i{4\pi}\int d^{2|2}\tilde z
\,\Sigma(\ln\Sigma -1)+c.c.
\label{EffTwistedPot}
\ee
This effective twisted potential was found originally in
\cite{DAdda} using component field quantum computations. Here we
reproduced the same result using the method of covariant
perturbative computations in the $\cN=(2,2)$ superspace.

The above results can be readily
generalized to the case of electrodynamics with $N$ chiral
flavors $Q_i$, $i=1,\ldots,N$, with charges $q_i$, see Eq.\
(\ref{SQi}). For the odd part of the effective action we have the
following modification of formulas (\ref{G-div}) and
(\ref{EffTwistedPot}):
\bea
\Gamma_{\rm div} &=& \frac{i}{8\pi\epsilon} \sum_{i=1}^N q_i \int
d^{2|2}\tilde z \,\Sigma + c.c.\,,\label{Gi-div}\\
\Gamma_{\rm fin} &=& -\frac i{4\pi}\sum_{i=1}^N q_i \ln q_i
 \int d^{2|2}\tilde z\, \Sigma
-\frac i{4\pi} \sum_{i=1}^N  q_i \int d^{2|2}\tilde z\,
 \Sigma(\ln\Sigma - 1)+c.c.\label{Gi-fin}
\eea
The equation (\ref{Gi-div}) implies that there is no infinite
renormalization of the FI parameter when the condition
(\ref{charge-zero}) is satisfied. This is the necessary condition for
existence of the Coulomb branch. This condition is also
sufficient for vanishing of the effective twisted potential in
(\ref{Gi-fin}). However, even when the condition
(\ref{charge-zero}) is satisfied, the first term in the right-hand side of (\ref{Gi-fin})
remains non-vanishing and yields a finite shift of the complex FI
parameter
\be
t\to t' = t+ \frac1{2\pi}\sum_{i=1}^N q_i \ln q_i\,.
\label{t-compl-shift}
\ee
The main effect of this finite quantum contribution is the shift
of the imaginary part $\theta$ of the complex FI parameter
(\ref{t-complex}). To compensate this shift, one has to add the corresponding
value to the classical FI parameter
\be
t =- \frac{\mathfrak m}{e^2} \langle {\rm Re}\,\sigma \rangle
- \frac1{2\pi}\sum_{i=1}^N q_i \ln q_i\,.
\label{t-special}
\ee
This is the sufficient condition of existence of the Coulomb branch on
the quantum level \cite{Hori}.

To summarize, we have shown that the FI parameter in the model
(\ref{S1}) receives infinite one-loop quantum contributions
(\ref{t-inf}) which lift the classical Coulomb branch. Such
infinite contributions may cancel among each other in the
generalization of the model (\ref{S1}) which involves $N$ charged
chiral flavors (\ref{SQi}). This happens when all charges of
chiral multiplets sum to zero (\ref{charge-zero}). However, there
is still a finite shift of the imaginary part of the FI parameter
as in (\ref{t-compl-shift}). Therefore, quantum Coulomb branch
exists when the classical FI parameter is tuned to a special value
(\ref{t-special}). Since we are interested in the effective
action in the Coulomb branch, in subsequent sections we will always
assume that the condition (\ref{t-special}) is satisfied.

We stress that all results of this subsection are not new; they
are well-known owing to \cite{DAdda,Witten93,Hori}. Here we just
summarized them for the sake of completeness of our
consideration.

\subsection{SQED with two chiral flavors}
\label{two-fl-sec}
\subsubsection{Classical action and background field setup}
Let us consider supersymmetric electrodynamics with two chiral
multiplets $Q_+$ and $Q_-$ carrying charges $\pm1$,
respectively,
\begin{subequations}
\label{S2all}
\bea
\label{S2}
S &=& S_{\rm gauge}[V] + S_{\rm mat}[Q,V]\,,\\
S_{\rm gauge}[V] &=& \frac1{2e^2} \int d^{2|4}z \,\Sigma\bar\Sigma
-\left[ \frac i2 \int d^{2|2}\tilde z \left( t
\Sigma + \frac{{\mathfrak m}}{2e^2}\Sigma^2\right)
+c.c.\right]\,,\label{S2b}\\
S_{\rm mat}[Q,V] &=& -\int d^{2|4} z ( \bar Q_+ e^{2 V} Q_+ + \bar Q_- e^{-2 V}
Q_-)
-\left(m\int d^{2|2}z \,Q_+ Q_-
+c.c.\right)\label{S2c}
,~~~~~~
\eea
\end{subequations}
where $m$ is the mass of the chiral multiplet while $\mathfrak
m$ is the vector multiplet mass. The latter is assumed to be small as
compared to the former,
\be
{\mathfrak m}^2\ll m^2 + \langle \Sigma\bar\Sigma\rangle\,.
\label{m<<m}
\ee
In this regime, we can study the effective action for the light
field $\Sigma$ which appears upon integrating out the heavy
chirals $Q_\pm$. In what follows, without loss of generality we will
assume that both $m$ and $\mathfrak m$ are real, though, in
general, they may be complex.

Let $\varphi_\pm$ be
scalar fields in the chiral multiplets
\be
\varphi_\pm = Q_\pm |\,,\qquad
\bar\varphi_\pm = \bar Q_\pm|\,.
\ee
After elimination of auxiliary fields, one can readily find the
scalar potential
\be
{\mathscr V} = \frac{e^2}{2}(\varphi_+\bar\varphi_+ -\varphi_-\bar\varphi_- -
t
 - \frac{\mathfrak m}{e^2} {\rm Re}\,\sigma)^2
+ (\sigma\bar\sigma+m^2 )(\varphi_+\bar\varphi_+ + \varphi_-\bar\varphi_-)
\,.
\ee
Similarly as in the model (\ref{S1}), the classical Coulomb branch
(\ref{ColBranch}) is possible at the special value of the FI parameter
(\ref{t-zero}). However, as is explained in the previous section,
the imaginary part of the FI parameter receives finite one-loop
contributions as in Eq.\ (\ref{t-compl-shift}). To compensate this
contribution, we have to set up the corresponding value to the
classical FI parameter
\be
t =\frac i 2
- \frac{\mathfrak m}{e^2} \langle {\rm Re}\,\sigma \rangle\,.
\label{t-const}
\ee
This allows us to study the low-energy effective action for the
light vector multiplet which appears by integrating out heavy
chiral multiplets beyond one-loop order.

In the framework of the background field method, we split the gauge superfield $V$
into background $V$ and quantum $v$ parts\footnote{The background gauge
superfield is denoted by the same letter as the original superfield $V$.
This should not lead to any confusions as the original unsplit gauge superfield
does not show up after the background-quantum splitting.}
\be
V \to V+ e\,v\,.
\ee
Upon this splitting, the actions
(\ref{S2}) and (\ref{S2c}) decompose as
\begin{subequations}
\bea
S_{\rm gauge}[V]&\to& S_{\rm gauge}[V]
+\frac1e \int d^{2|4}z\, v(iD^\alpha W_\alpha +2{\mathfrak m}\Sigma + 2e^2 t)
\nn\\&&
+\int d^{2|4} z \, v\left( \frac18 D^\alpha \bar D^2 D_\alpha
+i{\mathfrak m} D^\alpha\bar D_\alpha \right) v\,,
\label{label}\\
S_{\rm mat}[Q,V] &\to& S_{\rm mat}[{\cal Q},v]\,,
\eea
\end{subequations}
where ${\cal Q}_\pm$ and $\bar{\cal Q}_\pm$ are covariantly
(anti)chiral superfields with respect to the background gauge
superfield
\be
\bar{\cal Q}_+ = \bar Q_+ e^{2V}\,,\quad
{\cal Q}_+ = Q_+\,,\quad
\bar{\cal Q}_- = \bar Q_- e^{-2V}\,,\quad
{\cal Q}_- = Q_-\,.
\label{calQ}
\ee

The operators $D^\alpha \bar D^2 D_\alpha$ and $D^\alpha\bar D_\alpha$ in
(\ref{label}) are degenerate and require gauge fixing. The gauge
fixing is implemented by adding to the action (\ref{label}) the
following term
\be
S_{\rm gf} =  \int d^{2|4} z \,v\left[ -\frac1{16}\{
D^2 , \bar D^2 \}
+\frac{i\mathfrak m}{4}(D^2 + \bar D^2)
\right]v\,.
\label{Sgf}
\ee
This gauge-fixing action appears upon inserting the standard delta-functions
$\delta[f-i\bar D^2 v]\times \delta[\bar f - i D^2 v]$ into the functional
integral over ${\cal D}v$ and averaging them
with appropriate weight (see \cite{BPS-review} for details of this procedure
in the three-dimensional case).
After gauge fixing, we get the action for `quantum'
fields
\begin{subequations}
\label{Squant}
\bea
S_{\rm quant} &=& S_2 + S_{\rm int}\,,\\
S_2 &=& -\int d^{2|4}z\left[v(\square -H)v + \bar{\cal Q}_+{\cal Q}_+ + \bar{\cal Q}_- {\cal
Q}_-\right]
-\left( m \int d^{2|2}z \,{\cal Q}_+ {\cal Q}_- \right)\,,
\label{S-quant-2}
\\
S_{\rm int} &=& -2 \int d^{2|4}z \,
[e(\bar{\cal Q}_+ {\cal Q}_+ - \bar{\cal Q}_- {\cal Q}_-)v
+e^2(\bar{\cal Q}_+ {\cal Q}_ + + \bar{\cal Q}_- {\cal Q}_-)v^2]
+O(e^3)\,,~~~~~~~~
\label{S-quant-int}
\eea
\end{subequations}
where
\be
H=\frac{i{\mathfrak m}}{4}(2D^\alpha\bar D_\alpha + D^2 +\bar D^2)\,.
\label{H}
\ee
This operator obeys the important property
\be
H^2 = - {\mathfrak m}^2 \square\,.
\ee
This identity allows us to represent the propagator for the superfield
$v$ in the form
\bea
2i \langle v(z) v(z') \rangle &\equiv& G_v(z,z')=\frac1{\square-H}\delta^{2|4}(z-z')
\nn\\&=&-i \int_0^\infty ds
\left[
e^{-is{\mathfrak m}^2} + \frac{H}{{\mathfrak m}^2}
\left( 1-e^{-is {\mathfrak m}^2} \right)
\right] K_0(z,z'|s)\,,
\label{ph-prop}
\eea
where
\be
K_0(z,z'|s) = -\frac1{4\pi s}e^{-\frac{i\rho^2}{4s}}\zeta^2 \bar\zeta^2\,.
\label{K0}
\ee
Here $\rho^m$, $\zeta^\alpha$ and $\bar\zeta^\alpha$ are the
components of the supersymmetric interval (\ref{super-interval}).

In addition to the photon propagator (\ref{ph-prop}), the action
(\ref{S-quant-2}) yields the propagators of chiral superfields
\bea
i\langle {\cal Q}_+(z) {\cal Q}_-(z') \rangle &=& - m
G_+(z,z')\,,\nn\\
i\langle {\cal Q}_+(z) \bar{\cal Q}_+(z') \rangle &=& G_{+-}(z,z')
= G_{-+}(z',z)\,,\nn\\
i\langle \bar{\cal Q}_-(z) {\cal Q}_-(z') \rangle &=& G_{-+}(z,z')\,,
\eea
where Green's functions $G_+$ and $G_{+-}$ are defined by the
equations (\ref{G+def}) and (\ref{G+-def-a}), respectively.

Using the form of cubic and quatric interaction vertices for
quantum fields in (\ref{S-quant-int}), we deduce the
formal decomposition of the effective action up to two-loop order
\begin{subequations}
\bea
\label{General}
\Gamma &=& \Gamma^{(1)} + \Gamma^{(2)}\,,\\
\Gamma^{(1)} &=& i\Tr \ln(\square_+ + m^2) \,,\\
\Gamma^{(2)} &=& -2e^2 \int d^{2|4}z d^{2|4} z'
[G_{+-}(z,z') G_{+-}(z',z)
+m^2 G_+(z,z')G_-(z,z')]G_v(z,z')\,.~~~~~~~
\label{Gamma2loop}
\eea
\end{subequations}
Here $\Gamma^{(1)}$ is the one-loop effective action while
$\Gamma^{(2)}$ takes into account two-loop quantum corrections.
These quantum contributions will be calculated separately in
the subsequent sections.

\subsubsection{One-loop effective action}
\label{one-loopEA}
The computation of the one-loop effective action in the model
(\ref{S2all}) is very similar to the one for SQED with one chiral
flavor considered in Section \ref{sec-SQED1}. However, it has
some important features.

First of all, the effective action in the model (\ref{S2all})
possesses no odd part with respect to the reflection
$V\to - V$. As is demonstrated in Section \ref{odd-eff-action},
the odd contributions to the effective action cancel against each other in the model
where the charges of flavors sum to zero, (\ref{charge-zero}).
Thus, we have to focus only on
the even part of the one-loop effective action.

The computation of the even part of the effective action goes
along the same lines as in Section \ref{even-eff-action}.
Following these steps, one arrives at the expression (\ref{G-chiral}),
with two simple modifications: (i) The result (\ref{G-chiral}) should be
multiplied by 2 as we have contributions from two chiral flavors now;
(ii) the mass parameter $m$ should be inserted,
\be
\Gamma^{(1)}=-\frac i{8\pi}\int d^{2|2}z
\int_0^\infty ds\, W^2e^{-is(\Sigma\bar\Sigma+m^2)}\frac{\tanh(sf/2)}{sf/2} + c.c\,.
\label{G-chiral2}
\ee

It is an instructive exercise to rewrite the functional (\ref{G-chiral2}) in the
full superspace. We give the details of this procedure for the
chiral part of (\ref{G-chiral2}); the antichiral part can be
analyzed in the same way.

At the first step, we identically
rewrite the chiral part of (\ref{G-chiral2}) as the sum of two
terms
\bea
\Delta&\equiv&-\frac i{8\pi}\int d^{2|2}z
\int_0^\infty ds\, W^2e^{-is(\Sigma\bar\Sigma+m^2)}\frac{\tanh(sf/2)}{sf/2}
\nn\\&=&
-\frac i{8\pi}\int d^{2|2}z
\int_0^\infty ds\, W^2e^{-is(\Sigma\bar\Sigma+m^2)}
\nn\\&&
-\frac i{8\pi}\int d^{2|2}z
\int_0^\infty ds\,
\frac{W^2 \bar D^2 \bar W^2}{4f^2} e^{-is(\Sigma\bar\Sigma+m^2)}\left(\frac{\tanh(sf/2)}{sf/2}-1\right),
\label{333_}
\eea
where, in the last line, we have inserted the unity, $1=\frac1{4f^2} \bar D^2 \bar W^2
$. In this identity, the operator $\bar D^2$ can be used to
restore the full superspace measure due to (\ref{measure-id}).
Then, after evaluation of the proper-time integral in the second
line of (\ref{333_}), we have
\bea
\Delta&=&-\frac1{8\pi} \int d^{2|2}z\frac{W^2}{\Sigma\bar\Sigma+m^2}\nn\\
&&+\frac{i}{8\pi}\int d^{2|4}z \int_0^\infty ds\,
e^{-is(\Sigma\bar\Sigma + m^2)}
\frac{W^2 \bar W^2}{f^2}\left(\frac{\tanh(sf/2)}{sf/2}-1\right)\,.
\label{334}
\eea

Next, we have to restore the full superspace measure in the first
line of (\ref{334}) using the operators $\bar D_\alpha$ from
$W_\alpha = \bar D_\alpha G = i(\gamma^3)_\alpha^\beta \bar D_\beta
H$, see (\ref{W-G}). Making use of properties of the
superfield strengths (\ref{cov-lin}) and (\ref{W-G}), one can
prove the identity
\be
\int d^{2|4}z\,{\cal F}(X) =
-\int d^{2|2}z\, W^2 [(X-m^2){\cal F}''(X)
+{\cal F}'(X)]\,,
\label{335}
\ee
for some function ${\cal F}(X)$ and $X\equiv \Sigma\bar\Sigma+m^2$.
Comparing the right-hand side of (\ref{335}) with the first line
of (\ref{334}), one finds the following differential equation for
this function
\be
(X-m^2) {\cal F}''(X) + {\cal F}'(X) = \frac1X\,,
\ee
with the general solution
\be
{\cal F}(X) = c_1 + c_2 \ln(X-m^2)
+\frac12\ln^2\frac{X-m^2}{m^2}
+ {\rm Li}_2\left( -\frac{m^2}{X-m^2} \right)\,,
\label{336}
\ee
where $c_1$ and $c_2$ are arbitrary constants of integration. The
terms with these constants drop out upon integration
over the full superspace owing to the properties
(\ref{Sigma-def}) and (\ref{Sigma-tw-chiral}). The remaining two
terms in (\ref{336}) allow us to get the full-superspace representation
for the first term in (\ref{334})
\be
-\frac1{8\pi} \int d^{2|2}z\frac{W^2}{\Sigma\bar\Sigma+m^2}
=\frac1{8\pi} \int d^{2|4} z\left[
\ln \Sigma \ln \bar\Sigma + {\rm Li}_2\left(
-\frac{m^2}{\Sigma\bar\Sigma}
\right)
\right]\,.
\label{338}
\ee
Note that the last term in (\ref{338}) vanishes in the limit $m=0$
owing to the identity ${\rm Li}_2(0)=0$. In this limit, the
expression (\ref{338}) coincides with the non-holomorphic
potential in (\ref{G-full}).

Recall that we considered here the chiral part of
(\ref{G-chiral2}). It can be shown that the antichiral part gives
the same contribution as (\ref{334}), so that
$\Gamma^{(1)}=2\Delta$. Thus, substituting (\ref{338}) into
(\ref{334}), we end up with the representation for the
one-loop effective action in the full superspace
\bea
\Gamma^{(1)}&=&\frac1{4\pi} \int d^{2|4} z\left[
\ln \Sigma \ln \bar\Sigma + {\rm Li}_2\left(
-\frac{m^2}{\Sigma\bar\Sigma}
\right)
\right] \nn\\&&
+\frac{i}{4\pi}\int d^{2|4}z \int_0^\infty ds\,
e^{-is(\Sigma\bar\Sigma + m^2)}
\frac{W^2 \bar W^2}{f^2}\left(\frac{\tanh(sf/2)}{sf/2}-1\right)\,.
\label{one-loop-result}
\eea
The term in the first line here can be interpreted as the one-loop
quantum correction to the effective K\"ahler potential for the twisted chiral superfield
$\Sigma$
\be
K^{(1)} = \frac1{4\pi}
\ln \Sigma \ln \bar\Sigma + \frac1{4\pi}  {\rm Li}_2\left(
-\frac{m^2}{\Sigma\bar\Sigma}
\right)\,.
\label{K1-res}
\ee
The second line in (\ref{one-loop-result}) is responsible for
higher-derivative corrections in the one-loop
Euler-Heisenberg-type action.

\subsubsection{Two-loop effective K\"ahler potential}
\label{333}
In principle, starting from (\ref{Gamma2loop}) it is possible to determine two-loop quantum
corrections to the Euler-Heisenberg-type action.\footnote{Note that in the non-supersymmetric two-dimensional electrodynamics
the Euler-Heisenberg effective action was studied in
\cite{HMS} up to the two-loop order. We point out that beyond one loop this effective
action in the supersymmetric QED cannot be found by simple composition
of non-supersymmetric results in scalar and spinor
electrodynamics.} However, the form of the resulting expression
appears not very illuminating as it involves numerous proper-time
integrations and may have very limited applications. Therefore, in this section we restrict
ourself to studying two-loop quantum corrections only to the
effective K\"ahler potential for the twisted chiral superfield
$\Sigma$. To this aim, it is sufficient to consider the gauge superfield background
constrained by
\be
W_\alpha =0 \,,\qquad
\bar W_\alpha =0\,,
\label{W=0}
\ee
while superfields $\Sigma$ and $\bar\Sigma$ are constant and
non-vanishing. In this approximation, the heat kernels (\ref{K+}) and (\ref{K+-})
reduce to
\bea
K_{+}(z,z'|s)&\approx&\frac1{4\pi s}
e^{-is\Sigma\bar\Sigma}
e^{-i\frac{\rho^2}{4s}}
\zeta^2  I(z,z')\,,\\
K_{+-}(z,z'|s)&\approx& -\frac1{4\pi s}
e^{-is\Sigma\Bar\Sigma} e^{-i\frac{\rho^2}{4s}}I(z,z')\,.
\label{K+-approx}
\eea

The two-loop effective action (\ref{Gamma2loop}) is given by the
sum of two terms, which we denote by $\Gamma_{\rm A}$ and $\Gamma_{\rm B}$,
respectively,
\begin{subequations}
\label{GAB}
\bea
\Gamma^{(2)} &=& \Gamma_{\rm A} + \Gamma_{\rm B}\,,\\
\Gamma_{\rm A} &=& -2e^2 \int d^{2|4}z d^{2|4} z'
\,G_{+-}(z,z') G_{+-}(z',z) G_v(z,z')\,,
\label{GA1}\\
\Gamma_{\rm B} &=& -2e^2 m^2 \int d^{2|4}z d^{2|4} z'
\, G_+(z,z')G_-(z,z')G_v(z,z')\,.
\label{GB1}
\eea
\end{subequations}
These two terms correspond to the Feynman graphs of types A and B in
Fig.\ 1. They have slightly different structure and need to be
considered separately. Note that the two-loop graph of the
topology `8' vanishes identically and, thus, does not show up in
(\ref{GAB}).
\begin{figure}[t]
\begin{center}
\setlength{\unitlength}{1mm}
\begin{picture}(150,50)
\thicklines
\qbezier(35,25)(35,30.74)(30.6,35.6)
\qbezier(30.6,35.6)(25.75,40)(20,40)
\qbezier(20,40)(14.26,40)(9.39,35.6)
\qbezier(9.39,35.6)(5,30.74)(5,25)
\qbezier(5,25)(5,19.26)(9.39,14.39)
\qbezier(9.39,14.39)(14.26,10)(20,10)
\qbezier(20,10)(25.74,10)(30.6,14.39)
\qbezier(30.6,14.39)(35,19.26)(35,25)
\qbezier(75,25)(75,30.74)(70.6,35.6)
\qbezier(70.6,35.6)(65.75,40)(60,40)
\qbezier(60,40)(54.26,40)(49.39,35.6)
\qbezier(49.39,35.6)(45,30.74)(45,25)
\qbezier(45,25)(45,19.26)(49.39,14.39)
\qbezier(49.39,14.39)(54.26,10)(60,10)
\qbezier(60,10)(65.74,10)(70.6,14.39)
\qbezier(70.6,14.39)(75,19.26)(75,25)
\qbezier(135,25)(135,30.74)(130.6,35.6)
\qbezier(130.6,35.6)(125.75,40)(120,40)
\qbezier(120,40)(114.26,40)(109.39,35.6)
\qbezier(109.39,35.6)(105,30.74)(105,25)
\qbezier(105,25)(105,19.26)(109.39,14.39)
\qbezier(109.39,14.39)(114.26,10)(120,10)
\qbezier(120,10)(125.74,10)(130.6,14.39)
\qbezier(130.6,14.39)(135,19.26)(135,25)
\put(5,25){\circle*{2}}
\put(35,25){\circle*{2}}
\put(45,25){\circle*{2}}
\put(75,25){\circle*{2}}
\put(105,25){\circle*{2}}
\put(135,25){\circle*{2}}
\put(39,24){$+$}
\put(2,35){$Q_+$}
\put(32,35){$\bar Q_+$}
\put(42,35){$Q_-$}
\put(72,35){$\bar Q_-$}
\put(102,35){$Q_+$}
\put(132,35){$Q_-$}
\put(2,13){$\bar Q_+$}
\put(32,13){$Q_+$}
\put(42,13){$\bar Q_-$}
\put(72,13){$Q_-$}
\put(102,13){$\bar Q_+$}
\put(132,13){$\bar Q_-$}
\put(8,27){$v$}
\put(30,27){$v$}
\put(48,27){$v$}
\put(70,27){$v$}
\put(108,27){$v$}
\put(130,27){$v$}
\qbezier(5,25)(6,27)(7,25)
\qbezier(7,25)(8,23)(9,25)
\qbezier(9,25)(10,27)(11,25)
\qbezier(11,25)(12,23)(13,25)
\qbezier(13,25)(14,27)(15,25)
\qbezier(15,25)(16,23)(17,25)
\qbezier(17,25)(18,27)(19,25)
\qbezier(19,25)(20,23)(21,25)
\qbezier(21,25)(22,27)(23,25)
\qbezier(23,25)(24,23)(25,25)
\qbezier(25,25)(26,27)(27,25)
\qbezier(27,25)(28,23)(29,25)
\qbezier(29,25)(30,27)(31,25)
\qbezier(31,25)(32,23)(33,25)
\qbezier(33,25)(34,27)(35,25)
\qbezier(45,25)(46,27)(47,25)
\qbezier(47,25)(48,23)(49,25)
\qbezier(49,25)(50,27)(51,25)
\qbezier(51,25)(52,23)(53,25)
\qbezier(53,25)(54,27)(55,25)
\qbezier(55,25)(56,23)(57,25)
\qbezier(57,25)(58,27)(59,25)
\qbezier(59,25)(60,23)(61,25)
\qbezier(61,25)(62,27)(63,25)
\qbezier(63,25)(64,23)(65,25)
\qbezier(65,25)(66,27)(67,25)
\qbezier(67,25)(68,23)(69,25)
\qbezier(69,25)(70,27)(71,25)
\qbezier(71,25)(72,23)(73,25)
\qbezier(73,25)(74,27)(75,25)
\qbezier(105,25)(106,27)(107,25)
\qbezier(107,25)(108,23)(109,25)
\qbezier(109,25)(110,27)(111,25)
\qbezier(111,25)(112,23)(113,25)
\qbezier(113,25)(114,27)(115,25)
\qbezier(115,25)(116,23)(117,25)
\qbezier(117,25)(118,27)(119,25)
\qbezier(119,25)(120,23)(121,25)
\qbezier(121,25)(122,27)(123,25)
\qbezier(123,25)(124,23)(125,25)
\qbezier(125,25)(126,27)(127,25)
\qbezier(127,25)(128,23)(129,25)
\qbezier(129,25)(130,27)(131,25)
\qbezier(131,25)(132,23)(133,25)
\qbezier(133,25)(134,27)(135,25)
\put(34,3){Type A}
\put(114,3){Type B}
  \end{picture}
\end{center}
\caption[b]{Two-loop supergraphs in $\cN=(2,2)$ supersymmetric electrodynamics.}
\label{fig1}
\end{figure}
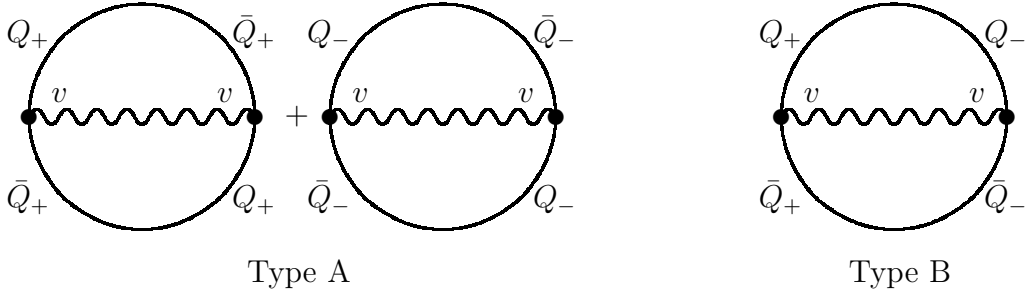

It is possible to show that the part of effective action
(\ref{GB1}) does not contribute to the effective K\"ahler
potential. In this formula, we express Green's functions via
the corresponding heat kernels (\ref{K+-def}) and (\ref{ph-prop})
\bea
\Gamma_{\rm B} &=& -2ie^2 m^2 \int d^{2|4}z d^{2|4}
z'\int_0^\infty ds\,dt\,du\,e^{-i(s+t)m^2}K_+(z,z'|s)K_-(z,z'|t)
\nn\\&&\times
\left[
e^{-iu{\mathfrak m}^2} + \frac{H}{{\mathfrak m}^2}
\left( 1-e^{-iu {\mathfrak m}^2} \right)
\right] K_0(z,z'|u)\,.
\eea
The operator $H$ in the last line contains the covariant spinor
derivatives (see Eq.\ (\ref{H})), which can be integrated by
parts
\bea
\Gamma_{\rm B} &=& 2ie^2 m^2 \int d^{2|4}z d^2\rho
\int_0^\infty \frac{ds\,dt\,du}{4\pi u}e^{-i(s+t)m^2} e^{-i\frac{\rho^2}{4u}}
\nn\\&&\times
\left[
e^{-iu{\mathfrak m}^2} + \frac{H}{{\mathfrak m}^2}
\left( 1-e^{-iu {\mathfrak m}^2} \right)
\right]K_+(z,z'|s)K_-(z,z'|t)\bigg|_{\zeta\to0}\,,
\label{GB2}
\eea
where we have taken into account the explicit form of the heat
kernel (\ref{K0}). The last line in (\ref{GB2}) contains the terms
of the following three types:
\begin{subequations}
\bea
&&K_+(z,z'|s)K_-(z,z'|t)\Big|_{\zeta\to0}\,,
\label{t1}\\
&&\nabla^2 K_+(z,z'|s) K_-(z,z'|t)\Big|_{\zeta\to0}\,,\qquad
K_+(z,z'|s)\bar \nabla^2 K_-(z,z'|t)\Big|_{\zeta\to0}\,,\label{t2}\\
&&\nabla^\alpha K_+(z,z'|s)\bar\nabla_\alpha
K_-(z,z'|t)\Big|_{\zeta\to0}\,.
\label{t3}
\eea
\end{subequations}
The terms involving (\ref{t1}) cannot contribute to the effective
K\"ahler potential since the expression (\ref{K+lim}) vanishes in
the approximation (\ref{W=0}). For the same reason there are no
contributions from the terms (\ref{t2}). In a similar way it is
easy to argue that the expressions (\ref{t3}) cannot contribute to the
effective K\"ahler potential owing to the properties
\be
\nabla_\alpha K_+(z,z'|s)|_{\zeta\to0} \propto W_\alpha\,,\qquad
\bar\nabla_\alpha K_-(z,z'|t)|_{\zeta\to0} \propto \bar W_\alpha\,.
\ee
These properties follow from the explicit form of the heat kernels
(\ref{K+}) and (\ref{K-}). Thus, in the approximation (\ref{W=0})
\be
\Gamma_{\rm B}=0\,.
\label{GB=0}
\ee

Now let us consider the contributions to the effective K\"ahler
potential from the part of the effective action (\ref{GA1}).
Making use of the identities (\ref{K2+-def}) and (\ref{ph-prop}),
this effective action can be cast to the form
\bea
\Gamma_{\rm A} &=&2ie^2 \int d^{2|4}z\, d^{2|4} z'\int_0^\infty
\frac{ds\,dt\,du}{4\pi u}e^{-i m^2(s+t)} e^{-\frac{i\rho^2}{4u}}
\nn\\&&\times
\left[
e^{-iu{\mathfrak m}^2} + \frac{H}{{\mathfrak m}^2}
\left( 1-e^{-iu {\mathfrak m}^2} \right)
\right] K_{+-}(z,z'|s)K_{+-}(z',z|t)\bigg|_{\zeta\to0}\,.
\label{240}
\eea
It is easy to argue that the terms with the operator $H$ in the
last line of (\ref{240}) give no contributions to the effective
K\"ahler potential. Indeed, covariant spinor derivatives in this
operator can hit the heat kernels yielding the terms
\bea
&&\nabla^2 K_{+-}(z,z'|s) K_{-+}(z,z'|t)\Big|_{\zeta\to0}\,,\qquad
K_{+-}(z,z'|s)\bar \nabla^2 K_{-+}(z,z'|t)\Big|_{\zeta\to0}\,,\nn\\
&&\nabla^\alpha K_{+-}(z,z'|s)\bar\nabla_\alpha
K_{-+}(z,z'|t)\Big|_{\zeta\to0}\,.
\label{tt3}
\eea
For such terms one can derive the following identities
\bea
\frac14\nabla^2 K_{+-}(z,z'|s)\Big|_{\zeta\to0} &=& i\frac{d}{ds} K_-(z,z'|s)\Big|_{\zeta\to0}
\propto \bar W^2\,,\nn\\
\frac14\bar\nabla^2 K_{-+}(z,z'|s)\Big|_{\zeta\to0} &=& i \frac{d}{ds} K_+(z,z'|s)\Big|_{\zeta\to0}
\propto W^2\,,\nn\\
\nabla_\alpha K_{+-}(z,z'|s)|_{\zeta\to 0} &\propto& \bar
W_\alpha\,,\qquad
\bar\nabla_\alpha K_{-+}(z,z'|t)|_{\zeta\to0} \propto W_\alpha\,.
\eea
Thus, contributions from these terms to the effective action
vanish in the approximation (\ref{W=0}).

Non-trivial contributions to (\ref{240}) appear only from the
terms without the operator $H$
\be
\Gamma_{\rm A} =\frac{ie^2}{32\pi^3} \int d^{2|4}z\, d^2\rho\int_0^\infty
\frac{ds\,dt\,du}{s\, t\, u}e^{-i (s+t)(m^2+ \Sigma\bar\Sigma)}
e^{-iu{\mathfrak m}^2} e^{-\frac{i\rho^2}{4}(s^{-1}+t^{-1}+u^{-1})}\,.
\label{3.82}
\ee
Here, the Gaussian integral over $d^2\rho$ can be easily evaluated
\be
\Gamma_{\rm A} = \frac{ie^2}{8\pi^2} \int d^{2|4}z \int_0^\infty\frac{ds\,dt\,du}{tu+su+st}
e^{-i u {\mathfrak m}^2}
e^{-i(s+t)(m^2 + \Sigma\bar\Sigma)}\,.
\label{3.83}
\ee
Finally, it is possible to perform integration over one of the
parameters, say $u$, and to represent the corresponding
contribution to the effective action in the form
\begin{subequations}
\bea
\Gamma_{\rm A} &=& \int d^{2|4}z\, K^{(2)}(\Sigma,\bar\Sigma)\,,
\\
K^{(2)}(\Sigma,\bar\Sigma)&=&\frac{ie^2}{8\pi^2}  \int_0^\infty \frac{ds\,dt}{s+t}
e^{-i(s+t)(m^2 + \Sigma\bar\Sigma)} e^{i\frac{{\mathfrak m}^2 s t}{s+t}}
E_1\left(\frac{i {\mathfrak m}^2 s\,t}{s+t}\right)\,,
\label{247}
\eea
\end{subequations}
where $E_1(z)$ is the exponential integral
\be
E_1(z) = \int_1^\infty \frac{dt}{t}e^{-t z}\,.
\ee

The expression (\ref{247}) represents the two-loop quantum
correction to the effective K\"ahler potential for the twisted
chiral superfield $\Sigma$. This formula involves integration over
the parameters $s$ and $t$ which are hard to evaluate for generic
values of masses $m$ and $\mathfrak m$. However, it is
possible to find explicitly the leading contributions to the
effective K\"ahler potential for small photon mass, i.e., in the
regime (\ref{m<<m}). In this case, we can use the asymptotics of
the function $E_1(ix)$ for small $x$,
\be
E_1(ix) = -\frac{i\pi}{2} - \gamma -\ln x + O(x)\,.
\ee
The integrals over $s$ and $t$ reduce to
\bea
\int_0^\infty \frac{ds\, dt}{s+t} e^{-(s+t)a} &=& \frac1a\,,\nn\\
\int_0^\infty \frac{ds\, dt}{s+t} e^{-(s+t)a } \ln\frac{s\,t}{s+t}
&=&
-\frac{2+\gamma}{a} - \frac1a \ln a\,,
\eea
where $a= i(m^2 + \Sigma\bar\Sigma)$. This yields the simple expression for (\ref{247})
\be
K^{(2)}(\Sigma,\bar\Sigma) = \frac{e^2}{4\pi^2}
\frac1{m^2+ \Sigma\bar\Sigma}
+\frac{e^2}{8\pi^2} \frac1{m^2 + \Sigma\bar\Sigma}\ln
\frac{m^2+\Sigma\bar\Sigma}{{\mathfrak m}^2}\,.
\label{K2-res}
\ee
This formula is a good approximation for the two-loop quantum
correction to the effective K\"ahler potential for $\Sigma$ in the
regime (\ref{m<<m}), i.e., when the photon possesses a small but
non-vanishing mass $\mathfrak m$. Obviously, (\ref{K2-res}) is
singular in the limit ${\mathfrak m}\to 0$. It emphasizes a
feature of two-dimensional electrodynamics that the quantum loop
diagrams with internal photon lines suffer from IR
singularities unless the photon possesses a mass.

In conclusion of this section, let us consider the full effective
K\"ahler potential $K(\Sigma,\bar\Sigma)$ which starts with the
classical value $\frac1{2e^2}\Sigma\bar\Sigma$ and includes both
one- and two-loop quantum corrections (\ref{K1-res}) and (\ref{K2-res})
\bea
K(\Sigma,\bar\Sigma) &=& \frac1{2e^2} \Sigma\bar\Sigma
+\frac1{4\pi}
\ln \Sigma \ln \bar\Sigma + \frac1{4\pi}  {\rm Li}_2\left(
-\frac{m^2}{\Sigma\bar\Sigma}
\right)
\nn\\&&
+\frac{e^2}{4\pi^2}
\frac1{m^2+ \Sigma\bar\Sigma}
+\frac{e^2}{8\pi^2} \frac1{m^2 + \Sigma\bar\Sigma}\ln
\frac{m^2+\Sigma\bar\Sigma}{{\mathfrak m}^2}\,.
\label{Kfin}
\eea
The corresponding sigma-model metric reads
\bea
ds^2 &=& \partial_z \partial_{\bar z} K(z,\bar z) dz
d\bar z \label{metric}\\
&=& \left(\frac1{2e^2}
+ \frac1{4\pi}\frac1{z\bar z +
m^2}
-\frac{e^2}{8\pi^2}\frac{m^2}{(m^2+z\bar z)^3}
+\frac{e^2}{8\pi^2}\frac{z\bar z - m^2}{(m^2+z\bar z)^3}
\ln
\frac{m^2+z\bar z}{{\mathfrak m}^2}
\right)dz d\bar z\,.\nn
\eea
For vanishing mass of the chiral multiplet, $m=0$, this metric
acquires a simple form
\be
ds^2|_{m=0}
= \left(\frac1{2e^2}
+ \frac1{4\pi}\frac1{z\bar z }
+\frac{e^2}{8\pi^2}\frac{1}{(z\bar z)^2}
\ln
\frac{z\bar z}{{\mathfrak m}^2}
\right)dz d\bar z\,.
\label{result-metric}
\ee
We stress that this metric makes sense for a small but non-vanishing photon mass
$\mathfrak m$.

The two-loop K\"ahler potential (\ref{Kfin}) and the corresponding
metric (\ref{metric}) are new results obtained here by direct
quantum computations in the $\cN=(2,2)$ superspace.
Though the one-loop quantum corrections to this metric were
found long ago in \cite{DAdda}, to the best of our knowledge the two-loop quantum
corrections have never been presented before.

\section{Low-energy effective action in $\cN=(4,4)$ SQED}
\setcounter{equation}{0}

\subsection{Classical action and loop expansion of the effective
action}
The $(4,4)$ vector multiplet may be described by the $\cN=(2,2)$
vector multiplet $V$ and a chiral multiplet $\Phi$. The
hypermultiplet is described by the pair of chiral fields $(Q_+,
Q_-)$. Let us consider the following action for these multiplets
\begin{subequations}
\label{S4}
\bea
S&=& S_{\rm gauge}[V,\Phi]  + S_{\rm
mat}[Q,V,\Phi]\,,\\
S_{\rm guage}[V,\Phi] &=& \frac1{2e^2} \int d^{2|4}z (\Sigma\bar\Sigma - \Phi\bar\Phi)
\nn\\&&
-\bigg[ \frac{ it}2 \int d^{2|2}\tilde z\,
\Sigma
+\frac{i{\mathfrak m}}{4e^2}
\left( \int d^{2|2} z\, \Phi^2 - \int d^{2|2}\tilde z\, \Sigma^2  \right)
+c.c.\bigg],\\
S_{\rm mat}[Q,V,\Phi] &=& -\int d^{2|4} z(\bar Q_+e^{2V}Q_+ + \bar Q_-e^{-2V} Q_-)
-\left(\int d^{2|2} z \, Q_+ \Phi Q_- +
c.c.\right).~~~
\eea
\end{subequations}
For ${\mathfrak m}=0$ this action is invariant under
`hidden' $(2,2)$ supersymmetry with anticommuting parameters
$\epsilon_\alpha$ and $\bar\epsilon_\alpha$
\bea
\delta V&=&\frac12(\bar\epsilon^\alpha \bar\theta_\alpha \Phi
 - \epsilon^\alpha \theta_\alpha \bar \Phi)\,,\nn\\
\delta \Phi&=&i\epsilon^\alpha W_\alpha\,,\qquad
 \delta\bar \Phi= i\bar\epsilon^\alpha \bar W_\alpha\,,\nn\\
\delta{\cal Q}_+&=&-\frac14 \bar\nabla^2(\bar\epsilon^\alpha \bar\theta_\alpha \bar
{\cal Q}_-)\,,\qquad
\delta{\cal Q}_-=\frac14 \bar\nabla^2(\bar\epsilon^\alpha \bar\theta_\alpha \bar
{\cal Q}_+)\,,\nn\\
\delta \bar{\cal Q}_+&=&-\frac14\nabla^2(\epsilon^\alpha\theta_\alpha
{\cal Q}_-)\,,\qquad
\delta \bar{\cal Q}_-=\frac14\nabla^2(\epsilon^\alpha\theta_\alpha
{\cal Q}_+)\,,
\label{hiddenSUSY}
\eea
where ${\cal Q}_\pm$ are as in (\ref{calQ}).
For non-vanishing photon mass, ${\mathfrak m}\ne0$, the action (\ref{S4})
is invariant under (\ref{hiddenSUSY}) only
for the real supersymmetry parameter $\bar\epsilon_\alpha =
\epsilon_\alpha$. This means that for generic ${\mathfrak m}$ the
model (\ref{S4}) describes the $\cN=(3,3)$ supersymmetric
electrodynamics while for ${\mathfrak m}=0$ the supersymmetry
extends up to $\cN=(4,4)$. This scenario is
completely analogous to the three-dimensional $\cN=4$
electrodynamics which can have only reduced $\cN=3$ supersymmetry
when the topological Chern-Simons mass term is turned on \cite{Kao1,Kao2}.
In our case, in (\ref{S4}) we keep non-vanishing photon mass $\mathfrak m$ in
order to get rid of IR singularities of Feynman graphs
beyond one loop. The one-loop contributions to the effective action, however, are independent of
$\mathfrak m$ and have the same form for both $\cN=(3,3)$ and
$\cN=(4,4)$ cases.

For quantizing the theory, we perform the background-quantum
splitting
\be
V \to V + e\, v\,,\qquad
\Phi \to \Phi + e\, \phi\,,
\ee
while the hypermultiplet $({\cal Q}_+, {\cal Q}_-)$ is considered
as the `quantum' superfield which will be integrated out in the
path integral. The background gauge superfield $V$ is constrained
by (\ref{W-on-shell}) and (\ref{cov-const}) while $\Phi$ is simply
constant
\be
D_\alpha \Phi = 0\,,\qquad
\bar D_\alpha \bar\Phi = 0\,.
\label{constPhi}
\ee

After adding the gauge fixing term (\ref{Sgf}), the `quantum' fields
are described by the action
\begin{subequations}
\label{Squant4}
\bea
S_{\rm quant}&=& S_2 + S_{\rm int}\,,\\
S_2 &=& -\int d^{2|4}z \left[v (\square-H) v + \frac 12 \phi\bar\phi+ {\cal Q}_+ \bar{\cal Q}_+
+ {\cal Q}_- \bar{\cal Q}_-\right]\nn\\&&
+\left[\int d^{2|2}z \left(\frac{i\mathfrak m}4\phi^2 -{\cal Q}_+ \Phi {\cal Q}_- \right)+ c.c.
\right],
\label{Sq2}
\\
S_{\rm int} &=& -2 \int d^{2|4}z [e ({\cal Q}_+ \bar {\cal Q}_+ - {\cal Q}_- \bar{\cal Q}_-)v
 + e^2({\cal Q}_+ \bar{\cal Q}_+ + {\cal Q}_- \bar{\cal
 Q}_-)v^2]\nn\\
 &&-e\int d^{2|2}z \, {\cal Q}_+ \phi {\cal Q}_-
 + e \int d^{2|2}\bar z\, \bar{\cal Q}_+ \bar\phi\bar{\cal Q}_- +
 O(e^3)\,.
\label{Sq3}
\eea
\end{subequations}
As compared with (\ref{Squant}), in (\ref{Squant4}) there are two
essential modifications: (i) in (\ref{Sq2}) we have the background
chiral superfield $\Phi$ in place of the mass $m$; (ii) in the last
line in (\ref{Sq3}) there are two additional vertices with the
quantum chiral superfield $\phi$ and its conjugate $\bar\phi$.
Taking these features into account, one can readily generalize
the $\cN=(2,2)$ effective action
(\ref{General}) to the $\cN=(4,4)$ (or, rather, $\cN=(3,3)$) case
\begin{subequations}
\label{4Gam-exp}
\bea
\Gamma &=& \Gamma^{(1)} + \Gamma^{(2)}\,,\\
\Gamma^{(1)} &=& i\Tr \ln(\square_+ + \Phi\bar\Phi) \,,\label{1loop4}\\
\Gamma^{(2)} &=& \Gamma_{\rm A}+ \Gamma_{\rm B} + \Gamma_{\rm C}
+ \Gamma_{\rm D}\,,
\label{G4-2loop}
\eea
\end{subequations}
where
\begin{subequations}
\bea
\Gamma_{\rm A} &=& -2e^2 \int d^{2|4}z d^{2|4} z'
\,G_{+-}(z,z') G_{+-}(z',z) G_v(z,z')\,,
\label{4GA}\\
\Gamma_{\rm B} &=& -2e^2 \int d^{2|4}z d^{2|4} z'
\, \Phi \bar\Phi\, G_+(z,z')G_-(z,z')G_v(z,z')\,,
\label{4GB}\\
\Gamma_{\rm C} &=&
2e^2 \int d^{2|4}z d^{2|4}z' \, G_{+-}(z,z')
G_{+-}(z,z') G_0(z,z')\,,
\label{4GC}\\
\Gamma_{\rm D} &=& \frac{ie^2{\mathfrak m}}{4}
\int d^{2|2}z d^{2|2}z'\,\Phi^2 \, G_+(z,z') G_+(z',z) \bar D^2
G_0(z,z')\nn\\&&
+\frac{ie^2{\mathfrak m}}{4} \int d^{2|2}\bar z d^{2|2}\bar z'\,
\bar\Phi^2 \, G_-(z,z')G_-(z',z) D^2 G_0(z,z')\,.
\label{4GD}
\eea
\end{subequations}
Here $G_v(z,z')$ is given by (\ref{ph-prop}) while $G_0(z,z')$ is
simply
\be
G_0(z,z') = \frac1{\square+{\mathfrak m}^2} \delta^{2|4}(z-z')
=-i\int_0^\infty ds \, e^{-is{\mathfrak m}^2} K_0(z,z'|s)\,.
\label{G0-prop}
\ee
Below, we compute separately the one- and two-loop contributions to
the effective action (\ref{4Gam-exp}).

\subsection{One-loop effective action and the Wess-Zumino term}

Recall that we consider the approximation (\ref{constPhi}) which
means that we discard any derivatives of (anti)chiral superfield
$\Phi$ ($\bar\Phi$). In this case the procedure of computation of
the one-loop effective action (\ref{1loop4}) is exactly the same
as in Section \ref{one-loopEA} for the $\cN=(2,2)$ SQED. Thus, we
can readily generalize the result (\ref{G-chiral2}) to the case of
$\cN=(4,4)$ SQED
\be
\Gamma^{(1)}=-\frac i{8\pi}\int d^{2|2}z
\int_0^\infty ds\, W^2e^{-is(\Sigma\bar\Sigma+\Phi\bar\Phi)}\frac{\tanh(sf/2)}{sf/2} + c.c\,.
\label{G-chiral4}
\ee

The effective action (\ref{G-chiral4}) is represented as a
functional in (anti)chiral superspace. It is instructive to
rewrite it in the full $\cN=(2,2)$ superspace. Following the same
procedure as in Section \ref{one-loopEA}, we find
\bea
\Gamma^{(1)}&=&\frac1{4\pi} \int d^{2|4} z\left[
\ln \Sigma \ln \bar\Sigma + {\rm Li}_2\left(
-\frac{\Phi\bar\Phi}{\Sigma\bar\Sigma}
\right)
\right] \nn\\&&
+\frac{i}{4\pi}\int d^{2|4}z \int_0^\infty ds\,
e^{-is(\Sigma\bar\Sigma + \Phi\bar\Phi)}
\frac{W^2 \bar W^2}{f^2}\left(\frac{\tanh(sf/2)}{sf/2}-1\right)\,.
\label{one-loop-result4}
\eea
The terms in the first line in (\ref{one-loop-result4}) are
leading in the derivative expansion of the effective action while
the terms in the second line correspond to higher-derivative
corrections. The leading terms
\begin{subequations}
\label{Gleading}
\bea
\Gamma_{\rm leading}&=&\int d^{2|4} z \,
K^{(1)}(\Sigma,\bar\Sigma;\Phi,\bar\Phi)\,,\\
K^{(1)}(\Sigma,\bar\Sigma;\Phi,\bar\Phi)
&=&\frac1{4\pi} \left[
\ln \Sigma \ln \bar\Sigma + {\rm Li}_2\left(
-\frac{\Phi\bar\Phi}{\Sigma\bar\Sigma}
\right)
\right]
\label{Kleading}
\eea
\end{subequations}
deserve several comments.

First of all, we point out the similarity of the superfield
expression (\ref{Gleading}) with the low-energy effective action of four-dimensional
$\cN=4$ SYM theory in $\cN=2$ superspace which was constructed in
\cite{BuIv}. Indeed, (\ref{Kleading}) contains the term
$\ln\Sigma\ln\bar\Sigma$ which is analogous to the non-holomorphic
potential for $\cN=2$ 4d superfield strength while the other terms
are very similar to the hypermultiplet completion of the
non-holomorphic potential which was constructed in \cite{BuIv}.
Surprisingly, such terms in 2d and 4d cases are described by the same ${\rm Li}_2$
function and have very similar form although they are given in
very different superspaces and for different models. Recall that
the $\cN=4$ susy-complete effective action in 4d $\cN=4$ SYM
theory was derived originally in \cite{BuIv} by imposing
the requirement of invariance under hidden supersymmetry
while in subsequent works this effective action was found by
direct quantum computations in superspace \cite{BIP,BBP,BP2005}
(see also \cite{BIP-review} for a review). In our case, we
obtained (\ref{Gleading}) as the leading part of the one-loop
effective action in $\cN=(4,4)$ SQED although originally it was
found in \cite{Rocek91} as a susy completion of the
non-holomorphic potential (\ref{2}).

The mentioned above similarity of (\ref{Gleading}) with the
low-energy effective action in 4d $\cN=4$ SYM theory is even
deeper. As was demonstrated in \cite{Belyaev1,Belyaev2} (see also
\cite{Review} for a review), the structure of leading terms in the
low-energy effective action in $\cN=4$ SYM theory can be recovered
from the fact that it contains the Wess-Zumino term for scalar
fields. This Wess-Zumino term is known to appear in the low-energy theory
as a result of 't~Hooft anomaly matching for $SU(4)$ R-symmetry of $\cN=4$
SYM theory \cite{Int}. Surprisingly, the effective action (\ref{Gleading})
may be given exactly the same interpretation. Indeed, in
\cite{Rocek91} the action of the form (\ref{Gleading}) was
proposed as a superfield generalization of a two-dimensional
sigma-model with the Wess-Zumino term. In our case, the appearance
of this term in the low-energy effective action is well understood.
Classically, the $\cN=(4,4)$ electrodynamics (\ref{S4}) respects
the $SU(2)\times SU(2)$ symmetry which is the R-symmetry of
$\cN=(4,4)$ Poincar\'e superalgebra. However, because of 't~Hooft
anomaly, this symmetry cannot be realized explicitly in the
low-energy theory but is still the symmetry of the effective Lagrangian up
to full derivative terms.
Recall that the effective action (\ref{Gleading}) is obtained upon
integrating out the hypermultiplet $(Q_+,Q_-)$ which contains
chiral fermions with respect to the R-symmetry group. Thus, in the low-energy theory the Wess-Zumino
term must appear as a response to the change of the number of
chiral fermions since the total contribution to the anomaly should
be the same regardless of the energy scale. This is the essence of the
't~Hooft anomaly matching argument \cite{tHooft}.

Let us derive the Wess-Zumino term for scalar fields from the
superfield action (\ref{Gleading}). The scalars appear
in the component field expansion of $\Phi$ and $\Sigma$ as follows
\bea
\Phi &= &\phi
+i\theta^\alpha\bar\theta^\beta \gamma^m_{\alpha\beta} \partial_m\phi
+\frac14\theta^2\bar\theta^2\square\phi+\ldots\,,\\
\Sigma &=& \sigma - i \theta^\alpha\bar\theta^\beta
(\gamma_m)_{\alpha\beta}
 \varepsilon^{mn} \partial_n \sigma - \frac14 \theta^2
 \bar\theta^2 \square \sigma + \ldots\,,
\eea
where dots stand for other component fields. Substituting these
expressions into (\ref{Gleading}) and integrating over the Grassmann variables one readily finds in the component field
expansion the Wess-Zumino term for the scalar fields
\be
S_{\rm WZ} = \frac1{2\pi} \int d^2x \frac{\phi\bar\phi}{\phi\bar\phi+\sigma\bar\sigma}
\varepsilon^{mn} \partial_m \alpha \partial_n \beta\,,
\label{SWZ}
\ee
where $\alpha$ and $\beta$ are phases of the complex scalars
$\phi$ and $\sigma$
\be
\phi = |\phi| e^{i\alpha}\,,\qquad
\sigma = |\sigma| e^{i\beta}\,.
\ee

The action (\ref{SWZ}) is explicitly invariant under $U(1)\times
U(1)$ symmetry which shifts the phases $\alpha$ and
$\beta$. This symmetry is the subgroup of the full $SU(2)\times
SU(2)$ R-symmetry group of the theory. It is possible to show that
(\ref{SWZ}) is implicitly invariant under $SU(2)\times
SU(2)\simeq SO(4)$ since this is the symmetry of the Wess-Zumino term
modulo total derivative terms. To show this, let us introduce real
scalars $X_A = (X_1,X_2,X_3,X_4)$ which transform as a vector under $SO(4)$
\be
\phi = X_1 + i X_2\,,\qquad
\sigma = X_3 + i X_4\,.
\ee
Then, the action (\ref{SWZ}) can be rewritten in the form of
integral over a three-dimensional space $\Omega$ which has
standard 2d Minkowski space as its boundary, $\partial\Omega = {\mathbb R}^{1,1}$
(see e.g. \cite{Belyaev1} for details)
\be
S_{\rm WZ} = \frac1{6\pi} \int_\Omega d^3x \frac1{(X_A X_A)^2}
\varepsilon^{ABCD} \varepsilon^{mnp}
X_A \partial_m X_B \partial_n X_C \partial_p X_D\,,
\label{SWZ1}
\ee
where $\varepsilon^{ABCD}$ and $\varepsilon^{mnp}$ are
antisymmetric tensors. The Wess-Zumino term in the form
(\ref{SWZ1}) has explicit $SO(4)$ symmetry.

One can reverse the arguments: once we know that the
Wess-Zumino term (\ref{SWZ1}) appears in the low-energy effective
action of $\cN=(4,4)$ SQED, we can immediately find
(\ref{Gleading}) as its supersymmetric generalization. However,
performing perturbative quantum computations we uncover not only
the leading term (\ref{Gleading}) in the low-energy effective
action, but also higher-derivative corrections which are encoded in the
second line of (\ref{one-loop-result4}).

We point out once more that (\ref{SWZ}) is explicitly invariant
under $U(1)\times U(1)\simeq SO(2)\times SO(2)$ which is one of
the maximal subgroups of the full R-symmetry group $SO(4)$. However,
there are two more inequivalent maximal subgroups: $SO(3)\simeq
SU(2)$ and $SU(2)\times U(1)$. We speculate that these subgroups
may be made manifest in other superfield descriptions of the
$\cN=(4,4)$ gauge theory such as the harmonic superspace
\cite{HSS1,HSS2,HSS3,HSS4,HSS5}. Recall that in the 4d $\cN=4$ SYM
theory the careful account of all maximal subgroups of the $SU(4)$
R-symmetry group resulted in different but equivalent superfield
descriptions of the low-energy effective action
\cite{Belyaev1,Belyaev2,Review,BISZ}. It is tempting to develop
similar ideas for the low-energy effective action in 2d supersymmetric gauge
theories.

\subsection[Vanishing of two-loop corrections to generalized
K\"ahler potential]{Vanishing of two-loop corrections to generalized
K\"ahler \protect\\ potential}

In the previous section we computed one-loop effective action
which contains the term (\ref{Gleading}) as the leading part in
the derivative expansion. In \cite{Seib97} it was claimed that
this potential is non-renormalized by higher-loop quantum
corrections. This section aims to demonstrate explicitly that
two-loop quantum corrections to the generalized K\"ahler potential
(\ref{Kleading}) cancel among each other.

In the two-loop expansion of effective action (\ref{G4-2loop}),
the terms $\Gamma_{\rm A}$ and $\Gamma_{\rm B}$ can be represented
by Feynman graphs which have the same structure as those in the $\cN=(2,2)$
SQED given in Fig.\ 1. The terms $\Gamma_{\rm C}$ and $\Gamma_{\rm
D}$ are new since they involve the propagators for the
(anti)chiral superfield $\phi$ ($\bar\phi$). These terms are
represented by the Feynman graphs in Fig.\ 2. To find the contributions to the
effective action from these terms it is sufficient to consider
the vector multiplet background constrained by (\ref{W=0}) and
(\ref{constPhi}).
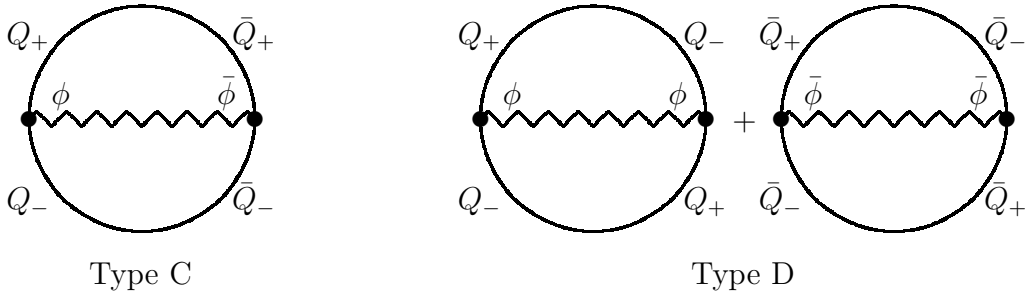
\begin{figure}[t]
\begin{center}
\setlength{\unitlength}{1mm}
\begin{picture}(150,50)
\thicklines
\qbezier(35,25)(35,30.74)(30.6,35.6)
\qbezier(30.6,35.6)(25.75,40)(20,40)
\qbezier(20,40)(14.26,40)(9.39,35.6)
\qbezier(9.39,35.6)(5,30.74)(5,25)
\qbezier(5,25)(5,19.26)(9.39,14.39)
\qbezier(9.39,14.39)(14.26,10)(20,10)
\qbezier(20,10)(25.74,10)(30.6,14.39)
\qbezier(30.6,14.39)(35,19.26)(35,25)
\qbezier(95,25)(95,30.74)(90.6,35.6)
\qbezier(90.6,35.6)(85.75,40)(80,40)
\qbezier(80,40)(74.26,40)(69.39,35.6)
\qbezier(69.39,35.6)(65,30.74)(65,25)
\qbezier(65,25)(65,19.26)(69.39,14.39)
\qbezier(69.39,14.39)(74.26,10)(80,10)
\qbezier(80,10)(85.74,10)(90.6,14.39)
\qbezier(90.6,14.39)(95,19.26)(95,25)
\qbezier(135,25)(135,30.74)(130.6,35.6)
\qbezier(130.6,35.6)(125.75,40)(120,40)
\qbezier(120,40)(114.26,40)(109.39,35.6)
\qbezier(109.39,35.6)(105,30.74)(105,25)
\qbezier(105,25)(105,19.26)(109.39,14.39)
\qbezier(109.39,14.39)(114.26,10)(120,10)
\qbezier(120,10)(125.74,10)(130.6,14.39)
\qbezier(130.6,14.39)(135,19.26)(135,25)
\put(5,25){\circle*{2}}
\put(35,25){\circle*{2}}
\put(65,25){\circle*{2}}
\put(95,25){\circle*{2}}
\put(105,25){\circle*{2}}
\put(135,25){\circle*{2}}
\put(2,35){$Q_+$}
\put(32,35){$\bar Q_+$}
\put(2,13){$Q_-$}
\put(32,13){$\bar Q_-$}
\put(8,27){$\phi$}
\put(30,27){$\bar\phi$}
\put(62,35){$Q_+$}
\put(92,35){$Q_-$}
\put(62,13){$Q_-$}
\put(92,13){$Q_+$}
\put(68,27){$\phi$}
\put(90,27){$\phi$}
\put(102,35){$\bar Q_+$}
\put(132,35){$\bar Q_-$}
\put(102,13){$\bar Q_-$}
\put(132,13){$\bar Q_+$}
\put(108,27){$\bar\phi$}
\put(130,27){$\bar\phi$}
\qbezier(5,25)(6,26)(6,26)
\qbezier(6,26)(7,25)(8,24)
\qbezier(8,24)(9,25)(10,26)
\qbezier(10,26)(11,25)(12,24)
\qbezier(12,24)(13,25)(14,26)
\qbezier(14,26)(15,25)(16,24)
\qbezier(16,24)(17,25)(18,26)
\qbezier(18,26)(19,25)(20,24)
\qbezier(20,24)(21,25)(22,26)
\qbezier(22,26)(23,25)(24,24)
\qbezier(24,24)(25,25)(26,26)
\qbezier(26,26)(27,25)(28,24)
\qbezier(28,24)(29,25)(30,26)
\qbezier(30,26)(31,25)(32,24)
\qbezier(32,24)(33,25)(34,26)
\qbezier(34,26)(35,25)(35,25)
\qbezier(65,25)(66,26)(66,26)
\qbezier(66,26)(67,25)(68,24)
\qbezier(68,24)(69,25)(70,26)
\qbezier(70,26)(71,25)(72,24)
\qbezier(72,24)(73,25)(74,26)
\qbezier(74,26)(75,25)(76,24)
\qbezier(76,24)(77,25)(78,26)
\qbezier(78,26)(79,25)(80,24)
\qbezier(80,24)(81,25)(82,26)
\qbezier(82,26)(83,25)(84,24)
\qbezier(84,24)(85,25)(86,26)
\qbezier(86,26)(87,25)(88,24)
\qbezier(88,24)(89,25)(90,26)
\qbezier(90,26)(91,25)(92,24)
\qbezier(92,24)(93,25)(94,26)
\qbezier(94,26)(95,25)(95,25)
\qbezier(105,25)(106,26)(106,26)
\qbezier(106,26)(107,25)(108,24)
\qbezier(108,24)(109,25)(110,26)
\qbezier(110,26)(111,25)(112,24)
\qbezier(112,24)(113,25)(114,26)
\qbezier(114,26)(115,25)(116,24)
\qbezier(116,24)(117,25)(118,26)
\qbezier(118,26)(119,25)(120,24)
\qbezier(120,24)(121,25)(122,26)
\qbezier(122,26)(123,25)(124,24)
\qbezier(124,24)(125,25)(126,26)
\qbezier(126,26)(127,25)(128,24)
\qbezier(128,24)(129,25)(130,26)
\qbezier(130,26)(131,25)(132,24)
\qbezier(132,24)(133,25)(134,26)
\qbezier(134,26)(135,25)(135,25)
\put(13,3){Type C}
\put(93,3){Type D}
\put(98.5,23.5){+}
  \end{picture}
\end{center}
\caption[b]{Two-loop supergraphs in $\cN=(4,4)$ SQED
which involve propagators $\langle \phi\bar\phi\rangle$, $\langle\phi\phi\rangle$
and $\langle\bar\phi\bar\phi\rangle$.}
\label{fig2}
\end{figure}

The details of computations of contributions to the effective action
(\ref{4GA}) and (\ref{4GB}) are exactly the same as those in
Section \ref{333}. We can immediately generalize the results
(\ref{GB=0}) and (\ref{3.83}) to the $\cN=(4,4)$ case
\bea
\Gamma_{\rm A} &=& \frac{ie^2}{8\pi^2} \int d^{2|4}z \int_0^\infty\frac{ds\,dt\,du}{tu+su+st}
e^{-i u {\mathfrak m}^2}
e^{-i(s+t)(\Phi\bar\Phi + \Sigma\bar\Sigma)}\,,
\label{4.19}\\
\Gamma_{\rm B} &=& 0\,.
\label{4.20}
\eea

It is easy to argue that the contribution to the effective action
(\ref{4GD}) vanishes. Indeed, the propagator $G_0$ (\ref{G0-prop})
contains the delta-function which implies that we need to consider
the heat kernels $K_+$ and $K_-$ at coincident points. As follows
from (\ref{K+lim}), $K_+(z,z'|s)\times K_+(z',z|t)|_{\zeta=0}=0$.
Thus, \be \Gamma_{\rm D} = 0\,. \label{4.21} \ee

It remains to consider the contribution $\Gamma_{\rm C}$ to the effective
action. Substituting here the propagators (\ref{K+-approx}) and
(\ref{G0-prop}) we have
\bea
\Gamma_{\rm C} &=& 2ie^2 \int d^{2|4}z d^{2|4}z' \int_0^\infty ds\,dt\,du\,
K_{+-}(z,z'|s)K_{+-}(z,z'|t) K_0(z,z'|u) e^{-i(s+t)\Phi\bar\Phi } e^{-iu{\mathfrak m}^2}
\nn\\&=&-\frac{2ie^2}{(4\pi)^3} \int d^{2|4}z d^2\rho \int_0^\infty
\frac{ds\,dt\,du}{s\,t\,u} e^{-i(s+t)(\Sigma\bar\Sigma+\Phi\bar\Phi)}
e^{-i\frac{\rho^2}4(s^{-1}+t^{-1}+u^{-1})}
e^{-iu{\mathfrak m}^2}\,.
\eea
After integration over $d^2\rho$ it becomes evident that this expression
contributes to the effective action with the opposite sign to (\ref{4.19})
\be
\Gamma_{\rm C}=-\frac{ie^2}{8\pi^2} \int d^{2|4}z \int_0^\infty
\frac{ds\,dt\,du}{st+su+tu} e^{-iu{\mathfrak m}^2} e^{-i(s+t)(\Sigma\bar\Sigma+\Phi\bar\Phi)}\,.
\label{4.23}
\ee
Thus, we conclude that the sum of the terms (\ref{4.19}), (\ref{4.20}),
(\ref{4.21}) and (\ref{4.23}) vanishes
\be
\Gamma^{(2)} = \Gamma_{\rm A}+\Gamma_{\rm B}+\Gamma_{\rm C}+\Gamma_{\rm
D}=0\,.
\ee
We stress that this does not mean that the complete two-loop
effective action vanishes, but just implies that there are no
two-loop quantum corrections to (\ref{Kleading}).

The non-renormalization of the generalized
K\"ahler potential $K(\Sigma,\bar\Sigma;\Phi,\bar\Phi)$ in (4,4)
gauge theories was claimed in \cite{Seib97}. In this section, we have explicitly
demonstrated the absence of two-loop quantum corrections to this
potential. There are also purely field-theoretical arguments that {\it
all} higher-loop quantum correction to this function vanish and
$K(\Sigma,\bar\Sigma;\Phi,\bar\Phi)$ is one-loop exact. Indeed, in
the previous section it was demonstrated that this potential is
responsible for the Wess-Zumino term for scalar fields
(\ref{SWZ}). It is well-known that appearance of Wess-Zumino terms
in low-energy effective action is strictly one-loop effect
associated with the 't~Hooft anomaly matching \cite{tHooft}. The
form of the Wess-Zumino term (\ref{SWZ1}) as well as the
coefficient in front of this action are rigidly fixed by
topological arguments (see e.g.\ \cite{Bardeen}). Therefore, the
function (\ref{Kleading}) cannot receive any higher-loop
corrections, and we end up with the exact result \be
K(\Sigma,\bar\Sigma;\Phi,\bar\Phi) = \frac1{2e^2}(\Sigma\bar\Sigma
- \Phi\bar\Phi) +\frac1{4\pi} \left[ \ln \Sigma \ln \bar\Sigma +
{\rm Li}_2\left( -\frac{\Phi\bar\Phi}{\Sigma\bar\Sigma} \right)
\right]\,. \label{Kexact} \ee The corresponding sigma-model metric
reads \be ds^2 = d\Sigma d\bar\Sigma \partial_\Sigma
\partial_{\bar\Sigma} K - d\Phi d\bar\Phi \partial_\Phi
\partial_{\bar\Phi}K =(d\Sigma d\bar\Sigma + d\Phi d\bar\Phi)
\left(\frac1{2e^2} + \frac1{\Sigma\bar\Sigma +
\Phi\bar\Phi}\right)\,. \ee As is demonstrated in \cite{GHR}, this
sigma-model corresponds to the generalized K\"ahler geometry with
$\cN=(4,4)$ extended supersymmetry. This geometry possesses
non-trivial torsion which can also be read off from
(\ref{Kleading}). Here we have demonstrated that this generalized
K\"ahler potential naturally arises as the leading term in the
low-energy effective action in $\cN=(4,4)$ gauge theory in the
Coulomb branch.

\section{Conclusions}

In this paper, we have studied two-loop quantum corrections to the low-energy
effective actions in the $\cN=(2,2)$ and $\cN=(4,4)$ SQED. In the
Coulomb branch, leading terms in the effective action in
$\cN=(2,2)$ SQED are represented by a superpotential and a
K\"ahler potential for superfield strengths described by a twisted
chiral superfield $\Sigma$. Although, at one-loop order, these
potentials were studied long ago \cite{DAdda}, to the best of our knowledge the two-loop quantum
corrections to the effective K\"ahler potential (\ref{K2-res}) have not been
presented before. The corresponding sigma-model metric in the
two-loop approximation is given by (\ref{result-metric}). We point
out that this metric depends on the vector multiplet mass $\mathfrak
m$ and is singular in the limit ${\mathfrak
m} \to 0$. This is a new feature of the two-dimensional case as
compared with the low-energy effective action of three-dimensional
\cite{3d4} and four-dimensional \cite{K1,K-Tyler} SQED where it was
well-defined for massless vector multiplet.

In the $\cN=(4,4)$ SQED, the leading part of the low-energy effective action
is described by the generalized K\"ahler potential $K(\Sigma,\bar\Sigma;\Phi,\bar\Phi)$
where $\Phi$ is a chiral and $\Sigma$ is a twisted chiral
superfields. We show that this potential is one-loop exact and is given
by (\ref{Kexact}). This potential was introduced for the first
time in \cite{Rocek91} in the study of two-dimensional sigma
models with torsion which originates from the Wess-Zumino term for
scalar fields. In our case, we demonstrate that the Wess-Zumino
term is the integral part of the low-energy effective action
associated with the 't~Hooft anomaly matching for the $SU(2)\times
SU(2)$ R-symmetry. The from of the potential $K$ appears
surprisingly similar to the low-energy effective action in 4d $\cN=4$
SYM theory obtained in \cite{BuIv}.

We have studied two-loop effective action in two-dimensional SQED
in the Coulomb branch. In is also very interesting to investigate
the structure of the effective action in the Higgs branch. Some of
the leading terms in this effective action were discussed in the
recent work \cite{Aharony} but the form of higher-loop quantum correction
is unknown. Another interesting problem is the study of quantum
aspects of two-dimensional gauge theories with semichiral
multiplets \cite{semichiral1,semichiral2,semichiral3}. Only some limited results in this direction
are available \cite{q-semichiral}, but the structure of
quantum corrections to the generalized K\"ahler potential remains
unknown.

\vspace{3mm}
{\bf Acknowledgments}\\[3mm]
I am very grateful to E.A. Ivanov for useful discussions.
I acknowledge the support from the RFBR grant No 15-02-06670.

\section{Appendix. $\cN=(2,2)$ superspace conventions}
\setcounter{equation}{0}
\renewcommand{\theequation}{A.\arabic{equation}}

In this paper, we use the two-dimensional $\cN=(2,2)$ superspace
which appears by the dimensional reduction from the
three-dimensional $\cN=2$ superspace or from the
four-dimensional standard $\cN=1$ superspace. Therefore, we employ the
superfield notation and conventions which are very close to the
ones used in the series of papers \cite{3d1,3d2,3d3,3d4,3d5} devoted to the study of
the low-energy effective actions in three-dimensional superfield
theories.

The 2d $\cN=(2,2)$ superspace is parametrized by the coordinates
$z^A=(x^m,\theta^\alpha, \bar\theta_\alpha)$, where
$x^m=(x^0,x^1)$ are the Minkowski space coordinates and
$\theta^\alpha=(\theta^1,\theta^2)$ are Grassmann coordinates
($\bar\theta_\alpha = (\theta_\alpha)^*$ are their complex
conjugate). The spinor indices are raised and lowered by means of
the antisymmetric $\varepsilon$-tensor,
$\theta^\alpha = \varepsilon^{\alpha\beta}\theta_\beta$,
$\theta_\alpha = \varepsilon_{\alpha\beta}\theta^\beta$,
$\varepsilon_{12}=\varepsilon^{21} = 1$.

One of the possible
choices for the two-dimensional gamma-matrices $(\gamma^m)=
(\gamma^m)_\alpha^\beta$ may be
$\gamma^0=-i\sigma_2$, $\gamma^1 = \sigma_1$, where $\sigma_i$ are
the Pauli matrices. The gamma-matrices obey the Clifford algebra
\be
\{ \gamma^m , \gamma^n \} = -2 \eta^{mn}{\bf 1}_{2\times 2}\,,\qquad
\eta^{mn} = {\rm diag} (1,-1)\,,
\ee
and possess the following orthogonality and completeness relations
\be
(\gamma^m)_\alpha^\beta (\gamma^n)^\alpha_\beta =
-2\eta^{mn}\,,\qquad
(\gamma^m)_\alpha^\beta (\gamma_m)^\rho_\sigma
=\varepsilon^{\beta\rho}\varepsilon_{\sigma\alpha}
-\delta_\alpha^\rho \delta^\beta_\sigma
+(\gamma^3)_\alpha^\beta (\gamma^3)^\rho_\sigma\,,
\ee
where $\gamma^3 = \gamma^1 \gamma^0$ is the chirality matrix. The chiral projectors are
$P_\pm =\frac12( {\bf 1}\pm \gamma^3)$.

The covariant spinor derivatives may be chosen in the form
\be
D_\alpha=\frac\partial{\partial\theta^\alpha}+i\bar\theta^\beta
(\gamma^m)_{\alpha\beta}\partial_m,\qquad
\bar D_\alpha=-\frac\partial{\partial\bar\theta^\alpha}
-i\theta^\beta (\gamma^m)_{\alpha\beta}\partial_m\,,
\label{A3}
\ee
with the following anti-commutation relation
\be
\{D_\alpha, \bar D_\beta
\}=-2i(\gamma^m)_{\alpha\beta}\partial_m\,.
\ee

The integration measure in the full $\cN=(2,2)$ superspace is
defined as
\be
d^{2|4}z \equiv d^2x d^4\theta = \frac 1{16} d^2x \, D^2 \bar
D^2\,,\quad\mbox{so that}\quad \int d^2x \, f(x) = \int d^{2|4} z
\,\theta^2 \bar\theta^2 f(x)\,,
\label{fullSSmeasure}
\ee
for some field $f(x)$. Here we adopt the following conventions for
contractions of spinor indices
\be
D^2 = D^\alpha D_\alpha\,,\quad
\bar D^2 = \bar D^\alpha \bar D_\alpha\,,\quad
\theta^2 = \theta^\alpha \theta_\alpha\,,\quad
\bar\theta^2 = \bar\theta^\alpha \bar\theta_\alpha\,.
\ee

The chiral and antichiral subspaces are parametrized by the coordinates $z_+ = (x_+^m ,
\theta_\alpha)$ and $z_- = (x_-^m,\bar\theta_\alpha)$, correspondingly,
 where
\be
x^m_\pm = x^m \pm i \gamma^m_{\alpha\beta} \theta^\alpha \bar\theta^\beta\,.
\ee
The integration measure in the
chiral subspace $d^{2|2}z \equiv d^2 x d^2\theta$ is related to the
full superspace measure (\ref{fullSSmeasure}) as
\be
d^{2|4}z = -\frac14 d^{2|2}z \bar D^2 = -\frac14 d^{2|2} \bar z
D^2\,,\quad\mbox{so that}\quad
\int d^2x \, f(x) = \int d^{2|2} z \,\theta^2 f(x_+)\,.
\label{measure-id}
\ee

Given a two-component spinor $\psi_\alpha$, we can consider each
of its components independently as they are Lorentz-invariant and
appear as the $P_\pm$ projections of $\psi$,
\be
\psi_\alpha \equiv (\psi_+,\psi_-)\,,\qquad
\psi_\pm = P_\pm \psi\,.
\ee
In a similar way, for a spinor with the upper spinor index,
$\chi^\alpha$, we have $\chi^\alpha \equiv(\chi^+ , \chi^-) = (-\chi_-,
\chi_+)$. As a consequence, it is possible to introduce twisted chiral coordinates
$\tilde z_+=(\tilde x^m_+, \theta_+, \bar\theta_-)$ and the
twisted antichiral ones $\tilde z_- = (\tilde x^m_-,
\theta_-,\bar\theta_+)$, where
\be
\tilde x^m_\pm = x^m \pm i(\gamma^m)_\alpha^\beta
\gamma^3_{\beta\gamma}\theta^\alpha \bar\theta^\gamma\,.
\ee
The integration measures over these subspaces are denoted by
$d^{2|2}\tilde z$ and $d^{2|2}\tilde{\bar z}$. They are related to the full superspace
measure (\ref{fullSSmeasure}) as
\be
d^{2|4}z = d^{2|2}\tilde z \frac12   D_+ \bar D_- = d^{2|2} \tilde{\bar
z}
\frac12  \bar D_+ D_-\,.
\ee
The existence of twisted chiral subspace in addition to the
conventional chiral one is the crucial feature of the
two-dimensional superspace as compared with the higher-dimensional
story.

\end{document}